\documentclass[10pt,a4paper]{article}%
\pdfoutput=1
\usepackage[latin1]{inputenc}
\usepackage{amsmath}
\usepackage{amsfonts}
\usepackage[usenames,dvipsnames]{color}
\usepackage{color}
\usepackage{graphicx} 
\usepackage{subfigure}
\newcommand{\be}{\begin{equation}}
\newcommand{\ee}{\end{equation}}
\newcommand{\bea}{\begin{eqnarray}}
\newcommand{\eea}{\end{eqnarray}}
\newcommand{\gphi}{\gamma_\varphi}
\newcommand{\glambda}{\gamma_\lambda}

\def\papertitlepage{\baselineskip 3.5ex\thispagestyle{empty}}
\def\preprinumber#1#2{\hfill\begin{minipage}{4.2cm} #1
        \par\noindent #2 \end{minipage}}
\allowdisplaybreaks[3]

\begin{document}

\papertitlepage
\setcounter{page}{0}
\preprinumber{KEK-TH-1375}{}
\baselineskip 0.8cm
\vspace*{2.0cm}

\begin{center}
{\Large\bf Antisymmetric field in string gas cosmology}
\end{center}
\vskip 4ex
\baselineskip 1.0 cm

\begin{center}

Igmar C. Rosas-L{\sc\'opez}$^{2)}$
\footnote{E-mail address: igmar@post.kek.jp}
and
Yoshihisa K{\sc itazawa}$^{1),2)}$
\footnote{E-mail address: kitazawa@post.kek.jp}\\
\vspace{5mm}
$^{1)}$
{\it KEK Theory Center}\\
{\it Tsukuba, Ibaraki 305-0801, Japan}\\
$^{2)}$
{\it The Graduate University for Advanced Studies (Sokendai)}\\
{\it Department of Particle and Nuclear Physics}\\
{\it Tsukuba, Ibaraki 305-0801, Japan}\\
\end{center}

\vskip 5ex
\baselineskip = 3.5 ex

\begin{center}{\bf Abstract}\end{center}

\hspace{0.7cm}
We study how the introduction of a 2-form field flux modify the dynamics of  a T-duality invariant string gas cosmology model
of Greene, Kabat and Marnerides. 
It induces a repulsive potential term in the effective action for the scale factor of the spacial dimensions. Without the 2-form field flux, the universe fails to expand when the pressure due to string modes vanishes. 
With the presence of a homogeneous 2-form field flux,  it propels 3 spacial dimensions to grow into a macroscopic 4 dimensional space-time. We find that it triggers an expansion of a universe away from the oscillating phase around the self-dual radius.
We also investigate the effects of a constant 2-form field. We can obtain an expanding 4 dimensional space-time
by tuning it at the critical value.

\vspace*{\fill}
\noindent
June 2010

\newpage

%%%%%%%%%%%
\section{Introduction}
\setcounter{equation}{0}

Since it was first proposed by Brandenberger and Vafa \cite{Brandenberger:1988aj,Brandenberger:2008nx,Tseytlin:1991xk,Tseytlin:1991ss}, the string gas cosmology scenario has generated a significant amount of interest. One of its most appealing characteristics is that it provides a mechanism for dynamically generating a four dimensional space-time. This argument is based on the assumption that strings interact mainly by intersecting each other. If that is the  case, the probability of intersection in space-time of two worldsheets has non-zero measure only if the dimension is equal or less than 4. %The mechanism, in simple terms is just that $2+2=4$. The worldsheets of the colliding strings are 2-dimensional, therefore, they should intersect in 4 dimensions. 
This is a classical argument and it is not obvious that it will remain true if quantum effects are taken into account. There have been several attempts at trying to formulate and prove the Brandenberger-Vafa mechanism with mixed results \cite{Battefeld:2005av,Sakellariadou:1995vk,Cleaver:1994bw,Bassett:2003ck,Easson:2001fy}. A recent work \cite{Greene:2009gp}, for example, succeeds in decompactifying 3 large spacial dimensions for a gas of diluted strings. 

Apart from the Brandenberger-Vafa mechanism, there have been other attempts to produce a mechanism for realizing a four dimensional space-time \cite{Kitazawa:1998, Kitazawa:2004}.  One of these scenarios consists in the inclusion of a  two-form field. This field is already present in the supergravity action,  hence, it is natural to consider its appearance in the equations of motion.
Cosmologies with a two-form field had been studied in the past and several solutions are known \cite{Copeland:1994km}. In the context of string gas cosmology, a two-form field flux was introduced for dilaton-gravity in \cite{Campos:2003ip, Campos:2005da}. In these solutions the 2-form field flux  is restricted to a 4-dimensional submanifold of space-time. The two-form field flux introduces a repulsive potential in the equations of motion for the scale factor of the spacial dimensions.  As a result the expansion of three spacial dimensions is enhanced and the corresponding scale factors become large.

In \cite{Greene:2008hf}, Greene et al. introduced a higher derivative dilaton gravity model.  This model replaces the Newtonian-like kinetic terms in the dilaton gravity action by their relativistic counterparts. By doing this, one obtains a model with some nice features: derivatives  with respect to the cosmic time become bounded, singularities at finite time are avoided, bounces on the scale factor are produced and loitering phases that solve the horizon problem are realized. Since this model respects T-duality, it has been applied to investigate stringy cosmology such as the Brandenberger-Vafa scenario. The results has shown that this model could lead to a  4 dimensional space-time only with a fine tuning. In general 3 large spacial dimensions are not preferred and any number of dimensions could become large.

Because of the many appealing features of this model,  it is interesting to investigate its behavior  in the presence of an antisymmetric tensor field. The effects of such kind of field have been studied in several works \cite{Copeland:1994km,Goldwirth:1993ha} in the context of dilaton gravity.  There have been some studies also for the string gas cosmology \cite{Campos:2003ip, Campos:2005da} case. For string gas cosmology, it is specially interesting to introduce a 2-form gauge field. Since the initial configuration is supposed to be compactified on a $d$-dimensional  torus, there exist  non-trivial effects even for a constant gauge field as strings  can wrap the compact dimensions.

In section 2, we briefly recall the low-energy string effective action. In section 3, we introduce a T-duality invariant effective action with 2-form field flux. We obtain analytic solutions of the effective action in several limiting cases. We find they can explain qualitative behaviors of the numerical solutions.
One of the main features of the string cosmology model is the introduction of a Hagedorn phase for the early universe. It  arguably removes the initial singularity  of the universe. 
We find that a homogeneous 2-form field flux triggers an expansion of a universe away from the oscillating phase around the self-dual radius.
An interesting stringy effect, as explained in section \ref{sec:constant_field}, is that an electric like two-form field  modifies the  effective string tension and the Hagedorn temperature \cite{Ambjorn:2000yr, Grignani:2001ik}. In such a case, the energy of the winding modes can vanish for the directions parallel to the electric field. We find that these spacial directions can expand even if the winding modes are present. We conclude in section 5.

%If we consider that the gas of strings has both strings oriented on the direction of the gauge field but also oriented in the opposite direction. For  the case of a  field near its critical value, the components with winding in the directing of the gauge field become light  and will dominate % need to check this asumption the partition function.
 % I think here we can suppose that a large number of light winding modes are produced in comparison with the number of modes produced with the opposite winding.
%Then, it is reasonable to consider, as we will do, that  there are only winding strings in the direction parallel to the gauge field. In the case of a magnetic field this peculiar situation.

\section{Low-energy string effective action}
\setcounter{equation}{0}
In this section we recall the low energy effective action for string theory. 
%This action can be given in different ways, depending on the conformal frame chosen. We present the equation of motion in the string frame and the Einstein frame.

%%%%%%%%%%%%
%\subsection{String frame}
The low energy effective action in the string frame is given by \cite{Lidsey:1999mc}
\be
S=\frac{1}{2\kappa_{10}^2}\int d^{10} x\sqrt{-g}e^{-\phi}
\left(R+(\nabla\phi)^2-V-\frac{1}{12}H^2\right)+\int d^{10}x\sqrt{-g}\mathcal{L}_{matter} \label{string-frame}
\ee
where $\kappa_{10}^2 =8\pi G_{10}$ and $H^2=H_{\mu\nu\lambda}H^{\mu\nu\lambda}$ with $H_{\mu\nu\lambda}=\partial_{[\mu}B_{\nu\lambda]}$. The sign convention is all $+$ according to the classification in Misner, Thorne and Wheeler \cite{WheelerGravitation}.
The variation of this action gives the equations of motion
\bea
&&R_\mu^\nu-\frac{1}{2}g^\nu_\mu R
=
\kappa_{10}^2e^\phi T_\mu^\nu+\frac{1}{12}(3H_{\mu\lambda\kappa}H^{\nu\lambda\kappa}-\frac{1}{2}g^\nu_\mu H^2)-\frac{1}{2}g_\mu^\nu V \nonumber\\
&&~~~~~~~~~~~~~~~~~~
-\frac{1}{2}g_\mu^\nu (\nabla\phi)^2 +(g_\mu^\nu g^{\lambda\kappa}-g_\mu^\lambda g^{\nu\kappa})\nabla_\lambda\nabla_\kappa\phi\\
&&\nabla_\mu (e^{-\phi}H^{\mu\nu\lambda})
= 0  \\
&&2\Box\phi +R- (\nabla\phi )^2 -V-\frac{1}{12}H^2 =0
\eea
where $T_{\mu\nu}$ is the energy-momentum tensor derived from the matter Lagrangian.
We assume the space-time metric is of the following type in the string frame
\be
ds^2=-dt^2+a^2dx_idx^i +b^2dx_Idx^I~,~~~i=\{ 1,2,3\}~, ~ I=\{ 4,...,9 \}
\ee
with
\be
a\equiv e^{\lambda (t)}~,\quad b\equiv e^{\nu (t)} \label{string-scalefac}
\ee

%%%%%%%%%%%%
 
%%%%%%%%%%%
%%%%%%%%%%%%%%
%%%%%%%%%%%%%%%%%%

%\subsection{Einstein frame}
By a conformal rescaling
\be
{\tilde{g}}_{\mu\nu}=e^{-\frac{\phi}{4}}g_{\mu\nu} \label{trans-string-einstein}
\ee
we obtain the effective action in the Einstein frame
\bea
S &=& \frac{1}{2\kappa_{10}^2}\int d^{10} x~\sqrt{-\tilde g}
\left(\tilde R-\frac{1}{8}(\tilde\nabla_\mu\phi)^2-V \text{e}^ \frac{\phi}{4}-\frac{1}{12}\text{e}^ {-\frac{\phi}{2}}\tilde H^2\right)
\nonumber\\ 
&&+\int d^{10}~x\sqrt{-\tilde g}~\text{e}^{-\frac{5\phi}{4}}\mathcal{L}_{matter}
\eea
The field equations  are 
\bea
\tilde R_{\mu\nu}-\frac{1}{2}\tilde g_{\mu\nu}\tilde R=\kappa_{10}^2 (\tilde T_{\mu\nu}+{}^{(H)}\tilde T_{\mu\nu}+{}^{(\phi )}\tilde T_{\mu\nu}+{}^{(V)}\tilde T_{\mu\nu})&& \\
\tilde\nabla_\mu (e^{-\frac{\phi}{2}}\tilde H^{\mu\nu\lambda})= 0&&  \label{eq:H}  \\
\tilde\Box\phi -Ve^{\frac{\phi}{4}}+\frac{1}{6}e^{-\frac{\phi}{4}}\tilde H^2 = 0&&
\eea
The homogeneous metric is given by
\be
d\tilde{s}^2=-d\tilde{t}^2+\tilde{a}^2dx_idx^i +\tilde{b}^2dx_Idx^I~,~~~i=\{ 1,2,3\}~, ~ I=\{ 4,...,9 \}
\ee
with
\be
\tilde{a}\equiv e^{\alpha(\tilde t)}~,\quad \tilde{b} \equiv e^{\beta (\tilde t)}~\label{Einstein-scalefac}
\ee
More detailed relations between the string frame and the Einstein frame are explained in the appendix A.
%This two conformal frames are equivalent and we can use any of them to study the cosmology. Even though in this work we mostly use the string frame for solving the equations of motion we will also use the Einstein frame in order to clarify the interpretation  of the results from the model.

We need a nontrivial solution for the field strength $H_{\mu\nu\lambda}$ to investigate its effects on the cosmology. The equation of motion for the two-form field (\ref{eq:H})
can be solved using  the Freund-Rubin ansatz \cite{Freund-Rubin}
\be
H^{\mu\nu\alpha} = e^{\phi}\epsilon^{\mu\nu\alpha\beta}\nabla_\beta  h ~,~\text{with}~~ \epsilon^{\mu\nu\lambda\kappa}=\frac{4!}{\sqrt{-g}}\delta^\mu_{[0}\delta^\nu_1\delta^\lambda_2\delta^\kappa_{3]}
\ee
We assume here that the two-form field flux exists only in three spacial directions. 
This assumption is consistent with the symmetry of the postulated space-time metric. 
Because $\nabla^\mu\epsilon_{\mu\nu\alpha\beta}=0$, the equation of motion is automatically satisfied. It only remains to satisfy  the closure condition \cite{Copeland:1994km} 
\be
\nabla_{[\beta} H_{\mu\nu\alpha ]}=0 
\ee
The equation of motion is then 
\be
\sum_{\mu ,\nu=0}^3 g^{\mu\nu}\partial_\mu ( a^3 b^{-6}e^{\phi} h,_\nu  )=0
\ee
In the case of a homogeneous field $h\equiv h(t)$, we obtain 
\be
 {\ddot h} +( 3\dot\lambda -6\dot\nu+\dot\phi) {\dot h}=0~~~~~
\ee
with $H_o$ a positive constant and $\dot{}\equiv \frac{d}{d t}$. This equation is solved by
\be
\dot h =\frac{\pm  H_0 { b}^6}{{ a}^3}e^{-\phi}
\ee
Thus, we have for these $ H_{\mu\nu\lambda}$
\bea
 H_{\mu\lambda\kappa} H^{\nu\lambda\kappa} &=& 0 ~~\text{if} ~~\mu~,\nu=\{0,4,5,...,9\} \\
 H_{i\lambda\kappa} H^{j\lambda\kappa} &=&\frac{2 H_o^2}{a^6}\delta^j_i  ~~\text{for} ~~i,~j=\{1, 2, 3\}
\eea
In particular we have $H^2=6 H_o^2a^{-6}$.

%%%%%%%%%%%%%
%%%%%%%%%%%%%%%%%
%%%%%%%%%%%%%%%%%%%%%%
%\subsection{Thermodynamics of the gas of strings}
%%%%%%%%%%%%%%%%%%
%%%%%%%%%%%%
%%%%%%%%%
As in \cite{Brandenberger:1988aj, Battefeld:2005av,Campos:2003ip,Greene:2008hf} we consider  a very simple setup with 3 types of matter: isotropic winding modes (with all winding numbers $W_i=W,~i=\{1,2,3\}$) with energies
\be
E_W=6We^\lambda ~ \label{energy-windingmodes}
\ee
isotropic momentum modes (with all momenta $K_i=K,~i=\{1,2,3\}$) with  energies
\be
E_K=6 K e^{-\lambda} 
\ee
and string oscillator modes that are modeled as pressureless dust with energy $E_\text{dust}$
\footnote{$E_\text{dust}$ also contains the contributions from strings with momenta and windings along 6 extra dimensions.}. 
The  total energy is the sum 
\be
E=E_W+E_K+E_\text{dust}+V
\ee
with $V= V(\varphi)$ the  potential for the dilaton. In an adiabatic system the pressures are
\bea
P_\varphi &=&\frac{\partial E}{\partial \varphi}=\frac{\partial F}{\partial\varphi}=-\frac{\partial L_\text{m}}{\partial \varphi}=\frac{\partial V}{\partial \varphi} ~\label{phi-pressure} \\
P_\lambda &=& -\frac{1}{3}\frac{\partial F}{\partial\lambda}=-\frac{1}{3}\frac{\partial L_\text{m}}{\partial\lambda}=-\frac{1}{3}\frac{\partial E}{\partial \lambda}=2K e^{-\lambda}-2We^\lambda ~\label{lambda-pressure}
\eea
The energy of the string gas is defined as $E_s \equiv E_W+ E_K+ E_\text{dust}$.

In order to model the behavior of the gas, we consider the following phases as in \cite{Greene:2008hf}:
\begin{itemize}
\item Hagedorn phase: thermal equilibrium at temperature $T_H=1/(\sqrt{8}\pi)$. The free energy of the gas vanishes ($P_\lambda =0$) and $E_s$ is conserved
\be
\langle  W\rangle  =\frac{\sqrt{E_s}}{12\sqrt{\pi}}e^{-\lambda} \quad ,\quad  
\langle  K\rangle  =\frac{\sqrt{E_s}}{12\sqrt{\pi}}e^{\lambda} 
\ee

\item Radiation phase: thermal equilibrium at $T<T_H$ with the universe dominated by  massless string modes.
In $d+1$ dimensional space-time, the internal energy is
\be
E_s=c_dV_dT^{d+1}, \qquad  c_d = 128\frac{2d!\zeta (d+1)}{(4\pi)^{d/2}\gamma(d/2)}(2-2^{-d})
\ee
with $V_d=(2\pi)^de^{d|\lambda |}$: the T-duality invariant volume.
\be
F=E_s-TS=-\frac{1}{d}c_dV_dT^{d+1}
\ee
\be
P_\lambda=\text{sign}(\lambda)E_s/d
\ee

\bea
\lambda >0: &&\langle W\rangle =0\quad ,\quad \langle K\rangle =\frac{1}{2}P_\lambda e^\lambda\quad \text{(radiation phase)} \\
\lambda <0: &&\langle W\rangle =-\frac{1}{2}P_\lambda e^{-\lambda}\quad ,\quad \langle K\rangle =0 \quad\text{(winding mode dominated phase)}
\eea
Note that the radiation and winding mode dominated phases are T-dual to each other.

\item Frozen phase: in this phase  the interactions between strings are turned off. The momentum and winding numbers are conserved, so $K$ and $W$ are frozen at the values they  have on Hagedorn exit. 
\item Non-equilibrium phase:
In order to model the string gas, we consider a phase in which the the temperature falls below the Hagedorn temperature. Since we also consider the interactions among the strings,  the expectation value of the momenta and winding number deviate from their equilibrium values such that the pressure of the string gas does not vanish. 

 \end{itemize}

%If we substitute the anzats
%\be
%\nabla_{[\mu}e^{2\phi}\epsilon_{\nu\lambda\kappa]\alpha}\partial^\alpha h =\frac{1}{4!}\epsilon_{\mu\nu\lambda\kappa}\epsilon^{\beta\gamma\delta\sigma}\epsilon_{\gamma\delta\sigma\alpha}\nabla_\beta (e^{2\phi}\partial^\alpha h)=\frac{1}{4}\epsilon_{\mu\nu\lambda\kappa}\nabla_\alpha (e^{2\phi}\partial^\alpha
%\ee h)
%where we have used 

%%%%%%%%%%%%%
%%%%%%%%%%%%%%%%%%
%%%%%%%%%%%%%%%%%%%%%%%%%
%%%%%%%%%%%%%%%%%%%%%%%%%%%%%%%%%%%%%%%%%%%%%%%%%%%%%%%%%%%%%%
%%%%%%%%%%%%%%%%%%%%%%%%%%%%%%%
\section{T-duality invariant action with two-form field }
\setcounter{equation}{0}
%%%%%%%%%%%%%%%%%%%%%%%%%%%%%%%%%%%%%%%%%%%%%%%%%%%%%%%
%%%%%%%%%%%%%%%%%%%
In order to analyze the effect of the two-form field, we work in the string frame. This allow us to choose a solution where the scale factor $\nu$, defined in (\ref{string-scalefac}), becomes constant and the analysis can be restricted to a 4-dimensional cosmology in the presence of a two-form field.

In the case of a homogeneous space-time \cite{Campos:2003ip, Campos:2005da, Goldwirth:1993ha}, the action (\ref{string-frame}) can be reduced to
\be
S=\int dt \big[ 4\pi e^{-\varphi}(d \dot\lambda^2 - \dot\varphi^2-U_0(\lambda)) + L_m \big]~ \label{originalS}
\ee
with $L_m$: the matter lagrangian, $\varphi $: related to the original dilaton $\phi$ as $\varphi =2\phi -d\lambda~.$  
The potential $U(\lambda)$ arises due to the nontrivial two-form field strength $H_{\mu\nu\lambda}$, as shown in \cite{Copeland:1994km,Campos:2003ip}
\be
H^2_{\mu\nu\alpha}=12 H_0^2 e^{-2 d \lambda}\equiv 24 U_0(\lambda)~ \label{Hmnl}
\ee
The parameter $d$ counts the number of spacial dimensions with the homogeneous scale factor $\lambda$.
Although our case corresponds to $d=3$ such that the space-time is 4-dimensional, we retain $d$ dependence explicitly in the equations of motion in order to keep track of the algebra. 

% We have to cite the paper where they give the  full solution
Now we proceed as \cite{Greene:2008hf} and replace the canonical kinetic terms by their higher derivative extensions. This leads to a phenomenological action with bounded velocities $\dot\varphi$, $\dot\lambda$. It thus rules out singularities at any finite proper time.  With this modification  we obtain a higher derivative action for the dilaton and scale factor which are coupled to a two-form field strength
\be
S=\int dt \Big[ 8\pi e^{-\varphi}\Big(  \sqrt{1- \dot\varphi^2} - \sqrt{1-d \dot\lambda^2} -U(\lambda)\Big) +L_m \Big]\label{action}~
\ee
$L_m=-F$ is the negative of the matter free energy (of the string gas) and $U(\lambda)$ is a  modified potential as explained below. 
%We present this potential and the justification for its introduction now.
%Note for me:  In this model we are supposing that some of the  dimensions freeze at the self-dual radius while the remaining ones evolve homogeneously. However, since the $B_{\mu\nu}$ field is defined on a 3 dimensional sub-manifold, if we consider the evolution of more than three dimensions there would be an inhomogeneous evolution between the coordinates affected directly by the form flux and the remaining ones.

%In this work, our main objective is to try to observe the effect that the form-flux has on the evolution of the scale factors directly affected by it, accordingly, we will neglect the evolution of additional coordinates and set $d=3$. This is equivalent to assume that that we have 7 spatial coordinates frozen at self-dual radius. 

String gas cosmology model needs to respect T-duality, a fundamental symmetry in string theory originating from the existence of the minimal length (string scale). It is realized as the symmetry between the winding and momentum modes in a toroidal compactification. Since (\ref{Hmnl}) is not explicitly invariant under T-duality, it is necessary to modify this potential in order to realize the symmetry. Such a modification allows us to solve the equations of motion near the self-dual radius numerically. An adequate choice is
\be
U(\lambda)\equiv\frac{1}{2}H_0^2(e^{2\lambda}+e^{-2\lambda})^{-d}=\frac{1}{2^{d+1}}H_0^2(\cosh 2\lambda)^{-d} \label{U}~
\ee
as (\ref{U}) is not singular at $\lambda=0$ and it reduces to (\ref{Hmnl}) for large $\lambda$.  

Defining the relativistic factors 
 \cite{Greene:2008hf}\be
\gamma_\varphi \equiv\frac{1}{\sqrt{1-\dot\varphi^2}}~,~~~~\gamma_\lambda \equiv\frac{1}{\sqrt{1-d\dot\lambda^2}}~
\ee
the equations of motion obtained from the action (\ref{action}) are
\bea
\dot\gamma_\varphi &=& \dot\varphi(\gamma_\varphi-\gamma_\lambda^{-1}-U(\lambda))+\frac{1}{8\pi^2}\dot\varphi e^\varphi P_\varphi~\label{eqphi} \label{eqphidot}\\
\dot\gamma_\lambda &=&\dot\varphi (\gamma_\lambda -\gamma_\lambda^{-1})-\dot\lambda\frac{\partial U}{\partial\lambda}+\frac{1}{8\pi^2}d\dot\lambda e^{\varphi}P_\lambda~ \label{eqlambda}\label{eqlambdadot}
\eea
We also need to impose the Hamiltonian constraint 
\be
\gamma_\varphi-\gamma_\lambda -U(\lambda)=\frac{1}{8\pi^2}Ee^\varphi~ \label{Friedmann}
\ee
where $E$ is the energy contained in matter. Notice that in the positive energy region $\gphi -\glambda >U(\lambda)$.
The pressure in (\ref{eqphi}), (\ref{eqlambda}) for the dilaton and the scale factor are defined in (\ref{phi-pressure}) and (\ref{lambda-pressure}). 
Rendering  equations  (\ref{eqphi}), (\ref{eqlambda}) into a  more manageable form, we obtain
\bea
\ddot\varphi &=&(1-\dot\varphi^2)\left[1-\gamma_\varphi^{-1}\big(\gamma_\lambda^{-1}+U(\lambda)-\frac{1}{8\pi^2}e^\varphi P_\varphi\big)\right] \label{eqdilaton}\\
\ddot\lambda &=&(1-d\dot\lambda^2)\left[\dot\varphi\dot\lambda -\gamma_\lambda^{-1}\Big(  \frac{1}{d}\frac{\partial U}{\partial \lambda}-\frac{1}{8\pi^2}e^\varphi P_\lambda  \Big)\right]~ \label{eqmotlambda}
\eea
%Substituting $U(\lambda)\equiv H_0^2(\cosh 2\lambda)^{-d}/2^{d+1}$, we find
%\bea
%\ddot\varphi &=& (1-\dot\varphi^2)\left[1-(1-\dot\varphi^2)^{1/2}\left((1-d\dot\lambda^2)^{1/2}-\frac{H_0^2}{2^{d+1}}(\cosh 2\lambda)^{-d}-\frac{1}{8\pi^2}e^\varphi P_\varphi\right)\right] \label{eqdilaton} \\
%\ddot\lambda &=&(1-d\dot\lambda^2)\left[\dot\varphi\dot\lambda +(1-d\dot\lambda^2)^{1/2}\Big(  \frac{H_0^2}{2^d}(\cosh 2\lambda)^{-d}\tanh 2\lambda+\frac{1}{8\pi^2}e^\varphi P_\lambda  \Big)\right]  \label{eqmotlambda}
%\eea

\begin{figure}
\begin{center}
\includegraphics[width=6cm,height=5cm]{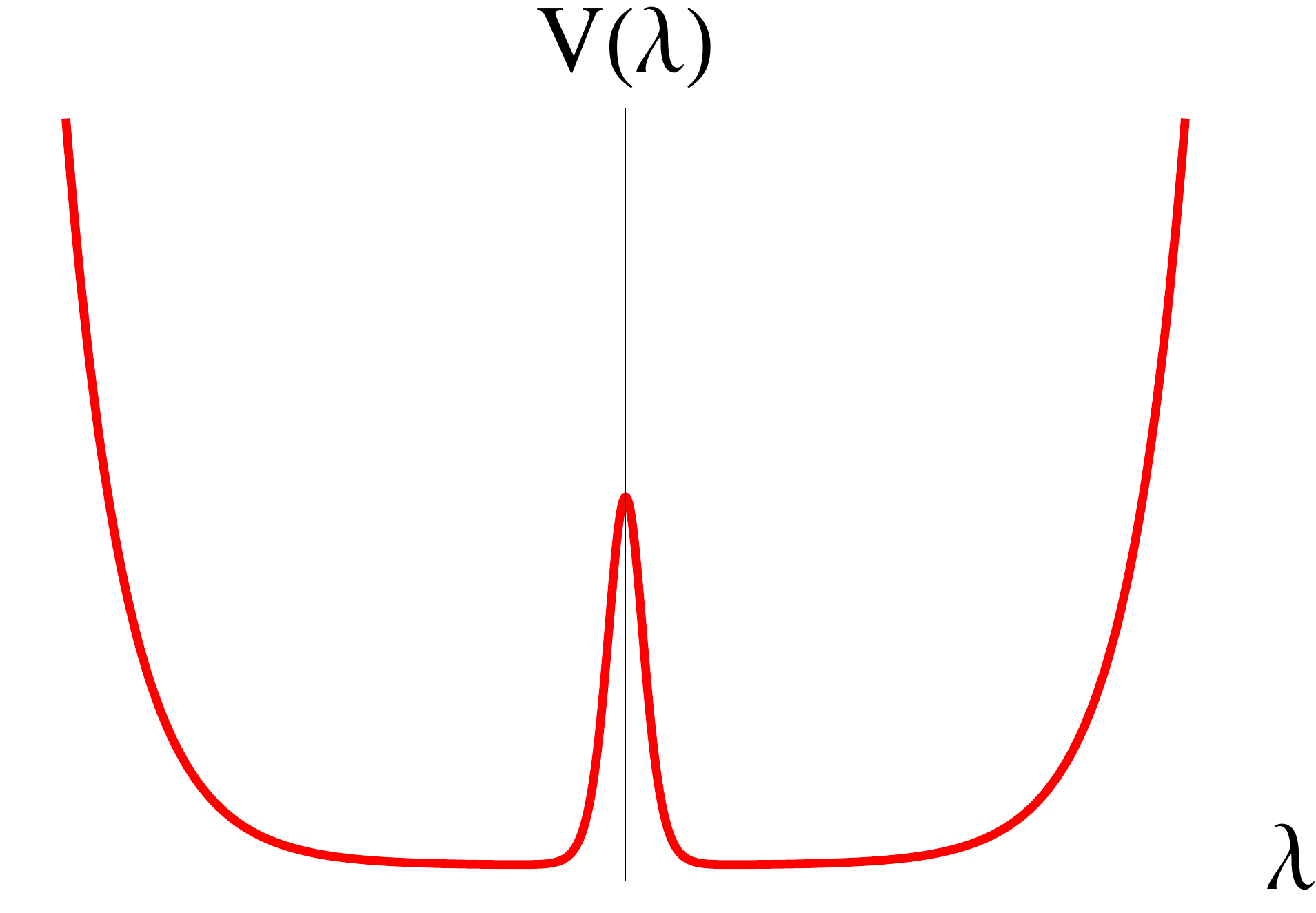}
\end{center}
\caption{Effective potential $V(\lambda )$ for the scale factor.}\label{Vlambda}
\end{figure}

%%%%%%%%%%%%%%%
%%%%%%%%%%%%%%%%%%
%%%%%%%%%%%%%%%%%%
%%%%%%%%%%%%%%%
Before trying to find some solutions to the equations of motion, let us examine the equation (\ref{Friedmann}) in order to get some idea of the expected behavior. If we put $U(\lambda)$ on the right side of the equation,  we see that $\lambda$ is subjected to the effective potential 
\be
V_\text{eff}(\lambda)=U(\lambda)+\frac{1}{8\pi^2}Ee^{\varphi}~\label{potential}
\ee
 A schematic plot of $V_\text{eff}(\lambda )$ is presented in figure \ref{Vlambda}.  If we assume, just for the moment, that the dilaton has some fixed value, we can observe the dependence of this potential on $\lambda$. The $E$ dependent term in (\ref{potential}) grows exponentially  as $\lambda $ increases if the winding  modes are present. Consequently, this term tends to  confine the scale factor near the self-dual radius.  On the other hand $U(\lambda )$ is a repulsive potential that has its maximun value at the self-dual radius. It decreases exponentially as  $\lambda $ increases. The term containing E, the energy of the string gas, is at the same time modulated by the exponential of the dilaton.  
 %In order to use the low energy approximation we suppose that the dilaton remains at weak coupling. 
 Then, as $\varphi \rightarrow -\infty$,  $V_\text{eff}(\lambda)$ flattens for large $\lambda$. As the confining effect of $V_\text{eff}(\lambda)$ diminishes in such a situation, $U(\lambda )$  becomes dominant and the scale factor is able to continue growing. It is also possible, depending on the initial conditions, for $\lambda$ to undergo oscillations  around one of the minima of the potential   or the self dual radius. In general, as the dilaton is going to weak coupling, these oscillations stop and the scale factor is forced to expand by $U(\lambda)$.

%Before proceeding to the  study  of the dynamics of the model,
%t that $H_{\mu\nu\lambda}$ has on the dynamics of the model.
%we should make some comments about the assumption  $d=3$ for the model. 
We have explained that the phenomenological action (\ref{action}) with $d=3$ is valid for a special class of solutions in superstring theory. In this paper, we investigate these solutions in which  only the scale factors for 3 spacial dimensions are time dependent. 
%We will verify this in the sections ahead where we will treat the case of anisotropic expansion of the universe.  % This approach can be argued to be plausible by the existence of solutions in which the scale factor remains at a constant value and since the field $H^{\mu\nu\lambda}$ drives the expansion of the 3 dimensional submanifold. 
The dynamics of this kind of cosmology has been studied before in several works \cite{Copeland:1994km, Goldwirth:1993ha,Lidsey:1999mc} where solutions for the case $d=3$ have been found. 
 
We should also be careful to point out that physical interpretation may depend on a chosen conformal frame.
%In  string gas cosmology,  the string frame has been prefered, but  in  an alternative way, we may work out the solutions in the Einstein frame. 
%Both conformal frames are equivalent and the solutions should contain the same physical information. From these equivalence, a problem for the interpretation of the model arises.
%This point can be observed in the following example: the vacuum solution when $t\rightarrow \infty$ has the following form in the string frame \cite{Easther:2003dd}
%\be
%e^\varphi\sim\frac{\text{const.}}{t^2}~,\quad  e^{\lambda}\sim \text{const.}
%\ee
%where $\varphi =\phi - d\lambda$ is the shifted dilaton. In this solution the scale factor becomes constant. If we now change variables to the Einstein frame,
% $\lambda =\alpha +\phi/(d+1)$, with $\alpha$ the scale factor in the Einstein frame, we find
%\be
%e^\alpha\sim (\text{const.})t^{\frac{2}{d+1}}~.
%\ee
%This implies that, when seen in the Einstein frame, the scale factor is explicitly time dependent and the universe is expanding. 
Unless we are able to fix the value of the dilaton $\phi$,  we can not consistently conclude that the size of a dimension  would remain small in the Einstein frame even if it becomes constant in the string frame.  %, if the scale factor in the string frame is constant that any scale factor stops expanding.
Nevertheless we may argue that the string frame is theoretically preferred to measure the size of the universe as T-duality holds
in the string frame.
Even with this limitation in mind, we will go ahead to study the cosmology in the string frame
\cite{Battefeld:2005av,Greene:2009gp,Campos:2003ip,Campos:2005da,Alexander:2000xv,Easther:2002mi, Easther:2003dd,Easther:2004sd}.
%even though we cannot identify the  physical scale factor  unless the dilaton is stabilized.

%%%%%%%%%%%%%%
%%%%%%%%%%%%%%%%%%%
%%%%%%%%%%%%%%%%%%%%%%%%
\subsection{Analytic solutions ($d=3$)}\label{Sec: SolutionsD4}
%%%%%%%%%%%%%%%%%%%%%%%%
%%%%%%%%%%%%%%%%%%%
%%%%%%%%%%%%%
Now we present some analytic solutions that can be obtained by solving the equations of motion. First, we assume a simple equation of state $P_\lambda=wE$, with $w$ a constant and $P_\varphi =0$ (no dilaton potential). Using (\ref{Friedmann}) and the equations of motion we get
\bea
\ddot\varphi &=& (1-\dot\varphi^2)\left[1-(1-\dot\varphi^2)^{1/2}\left((1-d\dot\lambda^2)^{1/2}-\frac{H_0^2}{2^{d+1}}(\cosh 2\lambda)^{-d}\right)\right] \label{w-eqmotphi}  \\
\ddot\lambda &=&(1-d\dot\lambda^2)\left[\dot\varphi\dot\lambda + \gamma_\lambda^{-1}\left(\frac{H_0^2}{2^d}(\cosh 2\lambda)^{-d}\tanh 2\lambda + w(\gamma_\varphi - \gamma_\lambda -\frac{H_0^2}{2^{d+1}}(\cosh 2\lambda)^{-d})\right)\right]~ \label{w-eqmotlambda} 
\eea
For the equation of state, we have three specific cases of interest: $w=0$, $w=1/d$ and $w=-1/ d$, that correspond to pressureless matter, radiation dominated era and winding mode dominated era respectively. 
As the boundary  condition for late time asymptotic behavior, we consider
\be
\dot\lambda\rightarrow 0~, ~~\dot\varphi\rightarrow 0~, ~~|\lambda |\rightarrow \infty~, ~~\varphi\rightarrow -\infty~ \label{latetimecond}
\ee
In this limit, the equations of motion (\ref{w-eqmotphi}), (\ref{w-eqmotlambda}) can be approximated as
\bea
\ddot\lambda &=& \dot\varphi\dot\lambda +H_0^2
e^{-2d\lambda} + \frac{w}{2} (\dot\varphi^2-d\dot\lambda^2-H_0^2e^{-2d\lambda}) \label{lambdaeqmot}\\
\ddot\varphi &=& \frac{1}{2}\dot\varphi^2 +\frac{1}{2}d\dot\lambda^2 +\frac{H_0^2}{2}e^{-2d\lambda} ~ \label{phieqmot}
\eea

%%%%%%%%%%%%
%%%%%%%%%%%%
%%%%%%%%%%%%
%%%%%%%%%%%%
\subsubsection{$H_0=0$, $w\neq 0$ case}
\begin{figure}[htp]
\begin{center}
\subfigure[]
{\includegraphics[width=7cm,height=6cm]{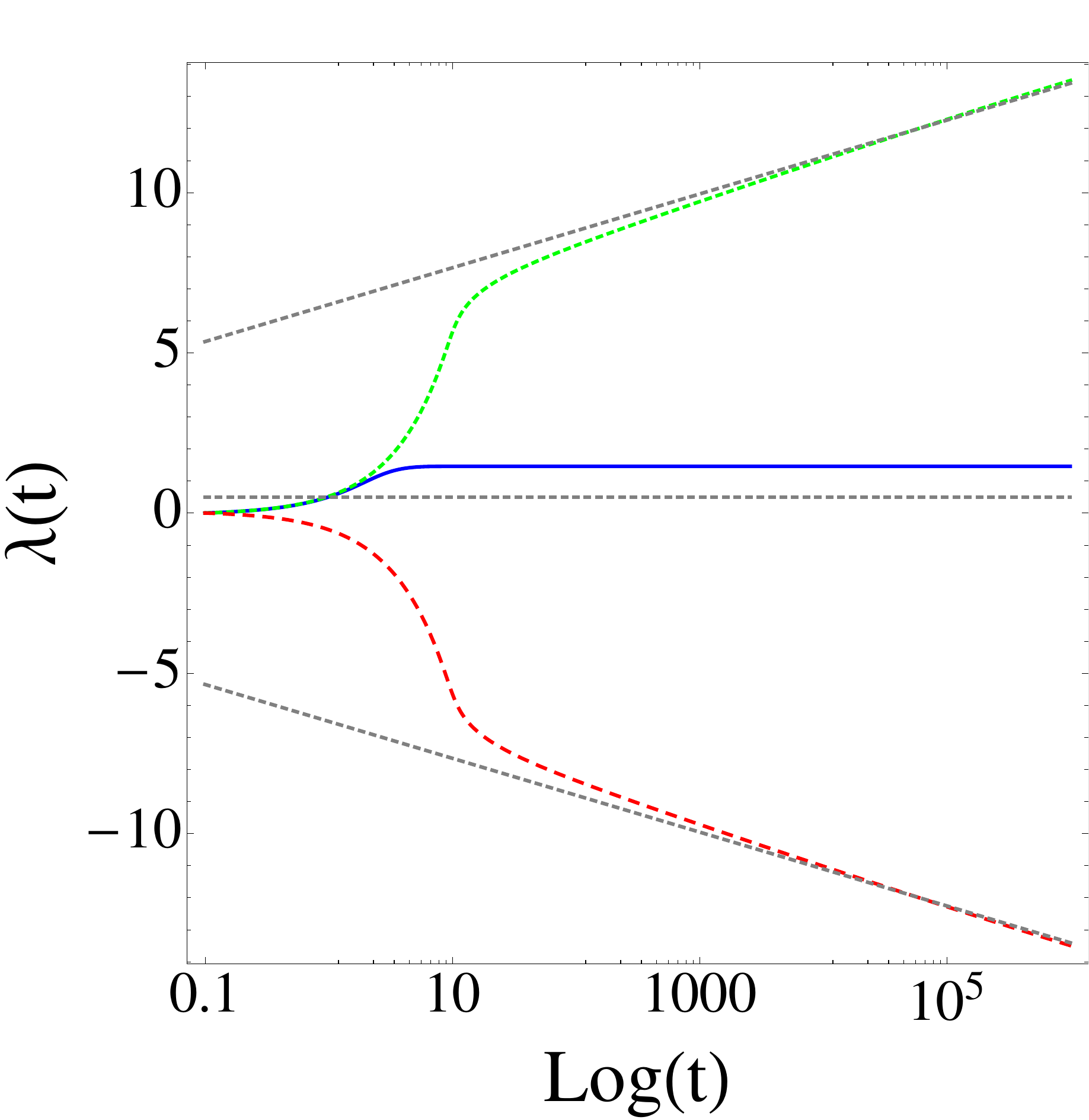}\label{lambda-radiation-dust-winding-a}} \quad
\subfigure[] 
{\includegraphics[width=7cm,height=6cm]{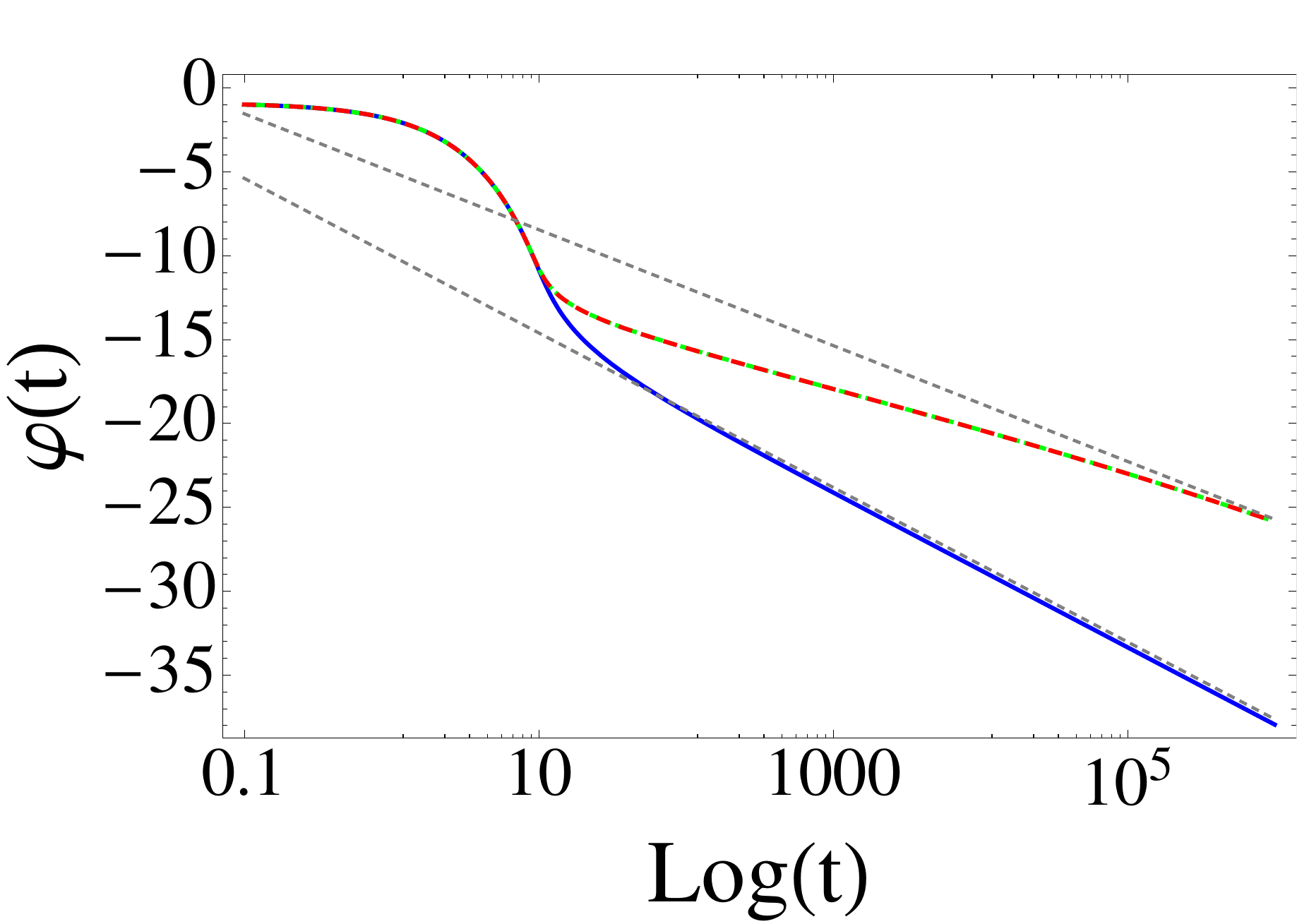}\label{lambda-radiation-dust-winding-b}}
\end{center}\label{lambda-radiation-dust-winding}
\caption{behavior of $\lambda$ and $\varphi$ with $H_0=0$ and {\color{Blue}$w=0$}, {\color{Green}$w=\frac{1}{3}$}, {\color{Red} $w=-\frac{1}{3}$}. The gray dotted lines correspond to the late time analytic solutions.}\label{radiation-dust-winding}
\end{figure}

%%%%%%%%%%%%%%
%%%%%%%%%%%%%%
We start with the standard string gas cosmology without 2-form field flux.
For  the case when $H_0=0$, we assume the following ansatz
\bea\label{anzats}
\varphi &=& A\log t +B~\label{anzatsphi}\\
\lambda &=& C\log t +D~\label{anzatslambda}
\eea
After substituting them in (\ref{lambdaeqmot}) and (\ref{phieqmot}), we find
\bea
\varphi &=&-\frac{2}{1+dw^2}\log t +B ~\label{phi-latetime-no-h}\\
\lambda &=& \frac{2w}{1+dw^2}\log t + D ~\label{lambda-latetime-no-h}
\eea
This asymptotic solutions can be seen  in figure \ref{radiation-dust-winding}  for $d=3$. 
%Here $H_o=0$ and $d$ can take any value. 
We have plotted  in the same figure the numerical solutions for the full equations of motion (\ref{w-eqmotphi}), (\ref{w-eqmotlambda})  with $H_o=0$: momentum mode dominated universe (green line, $w=1/3$), dust dominated universe (blue line, $w=0$) and  winding mode dominated universe (red line, $w=-1/3$). Of course the green and red lines are T dual to each other. The numerical solutions tend to the late time analytic solutions, which are plotted in figure \ref{radiation-dust-winding} as gray dotted lines.
%%%
%%%
%:figure: radiation, dust, windings
From this solution we note that the scale factor goes to a constant value very quickly in the absence of any driving pressure. This behavior can be seen also in figures \ref{radiation-dust-winding}, \ref{lambda-late-h},  \ref{momentum+winding+H}. The blue line in every picture represents the case when the effect of the pressure and the two-form field vanish,  leading to the  solution (\ref{phi-latetime-no-h}), (\ref{lambda-latetime-no-h}) with $w=0$.  Notice that this solution is valid for arbitrary $d$.
%In contrast, if we turn on the two-form field, after some time the term containing the $U(\lambda)$ potential becomes dominant and drives the expansion of the scale factor. Also, it alters the late time behavior of the shifted dilaton $\varphi$, as can be seen on figure \ref{lambda-late-b}, the slope of the log-linear plot is changed by the presence of the form flux.

For completeness, we mention that there is an additional solution  when $d=1$, $w=0$.  In this special case,  assuming $\dot\varphi =-\dot\lambda$, the equations of motion (\ref{lambdaeqmot}), (\ref{phieqmot}) reduce to a differential equation in one variable  $\ddot\varphi = \dot\varphi^2$, ($\ddot\lambda = -\dot\lambda^2  $).
Then we get the solution 
\bea
\varphi &=&-\log (t+A)+B \\
\lambda &=& \log (t+A) +C
\eea 
with $A$, $B$, $C$ constants. This solution is not physically relevant since we do not have the correct number of large space dimensions. Nevertheless, it is interesting to observe that a small coordinate can grow large even in the absence of any driving pressure.

%%%%%%%%%%%%%%%
%%%%%%%%%%%%%%%%%%
%%%%%%%%%%%%%%%%%%%%
\subsubsection{$H_0\neq 0$, $w= 0$ case}
\begin{figure}[htp]
\begin{center}
\subfigure[]
{\includegraphics[width=7cm,height=6cm]{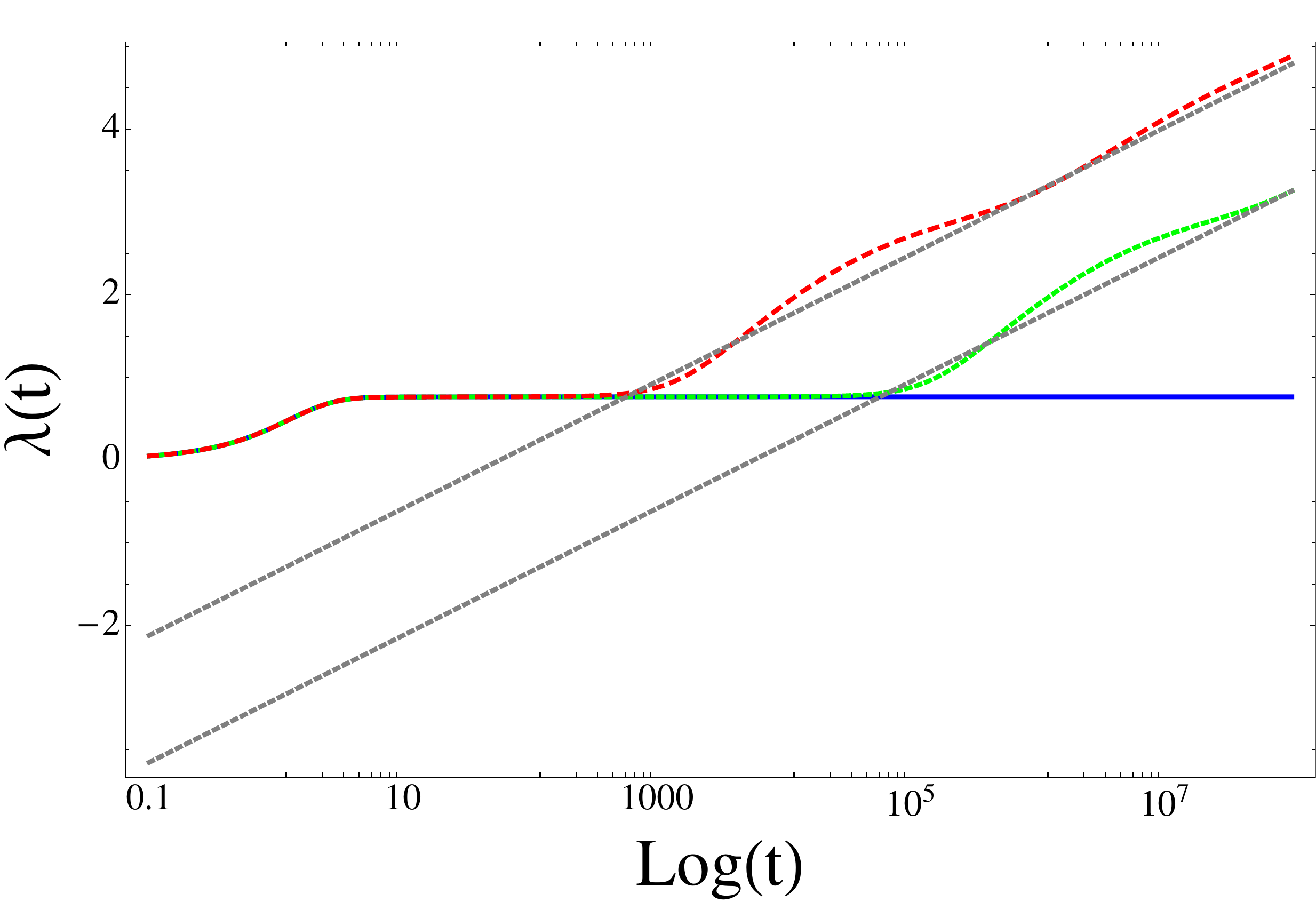}\label{lambda-late-h-a}} \quad
\subfigure[] 
{\includegraphics[width=7cm,height=6cm]{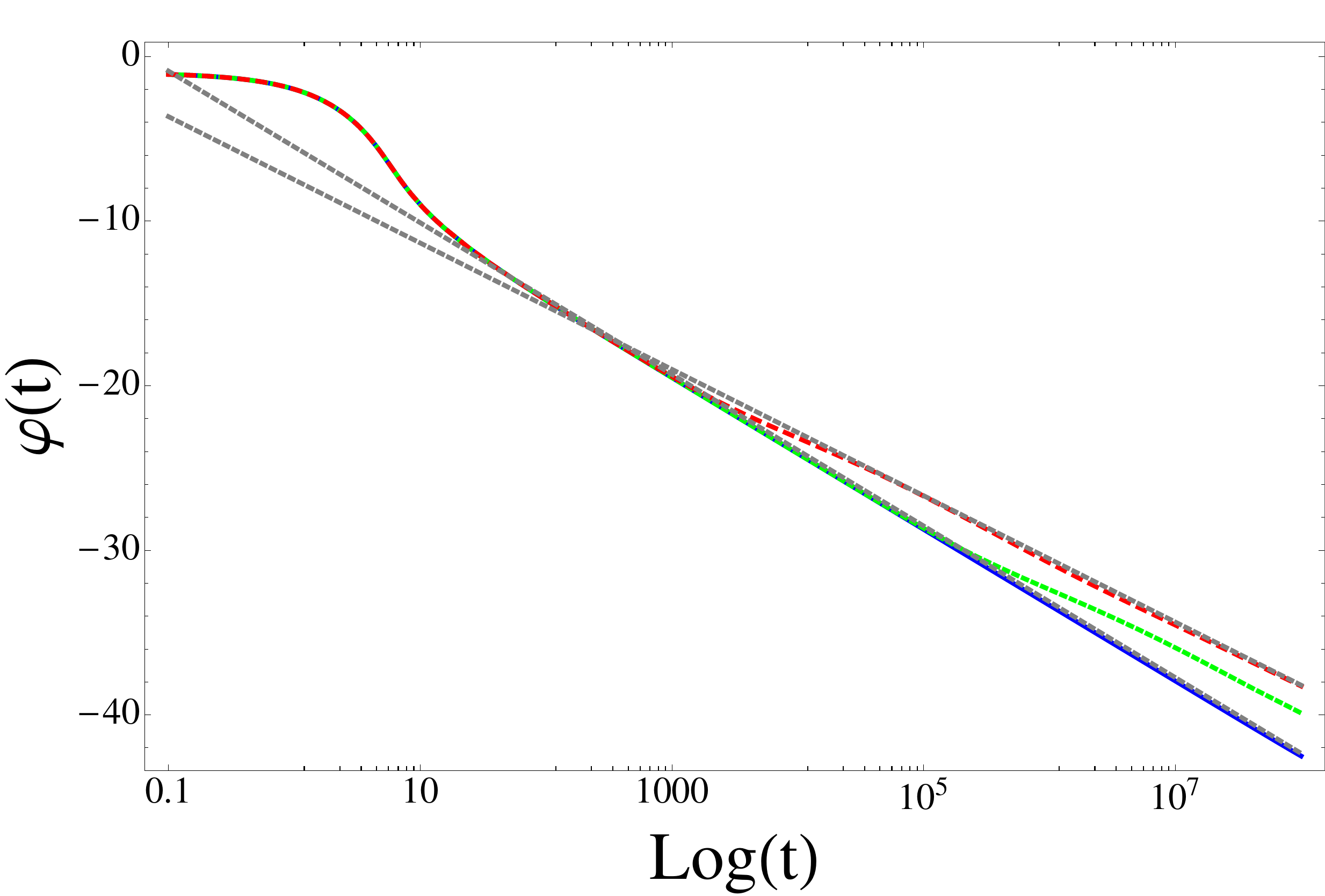}\label{lambda-late-h-b}}
\end{center}
\caption{behavior of $\lambda$ and $\varphi$ when there is only the two-form field flux present ($W=0$, $K=0$, {\color{Blue}$H_0=0$}, {\color{Red}$H_0=0.001$}, {\color{Green}$H_0=0.0001$} ).}\label{lambda-late-h}
\end{figure}

%%%%%%%%%%%%%%%%%%%%%
%%%%%%%%%%%%%%%%%%%
%%%%%%%%%%%%%%%%%
Now, we investigate a universe filled with dust ($w=0$) and an antisymmetric tensor potential ($H_0\neq 0$).  Under this conditions, we substitute the ansatz (\ref{anzats}) on equations (\ref{w-eqmotphi}), (\ref{w-eqmotlambda}) to leading  order as
\bea
\ddot\varphi&=&\frac{1}{2}\dot\varphi^2 +\frac{1}{2}d\dot\lambda^2 +\frac{H_0^2}{2}e^{-2d\lambda} ~ \label{H-pressurezero-phi}\\
\ddot\lambda &=&\dot\varphi\dot\lambda +H_0^2 e^{-2d\lambda} ~\label{H-pressurezero-lambda}
\eea
%In this case we have that $H_0$ is able, by itself, to induce the growth of a large scale factor, as can be seen in figure \ref{lambda-late-a}.  The fixed point associated with this behavior is 
%\be
%\dot\lambda\rightarrow 0~, ~~\dot\varphi\rightarrow 0~, ~~\lambda\rightarrow \infty~, ~~\varphi\rightarrow -\infty~. \label{latetimecond}
%\ee
Using ansatz (\ref{anzatsphi}), (\ref{anzatslambda}), we obtain 
%Combining both equations we can eliminate $H_0^2 e^{-2dD}$. Thus, we get  
 %\be 
%A=0  \quad \text{or\quad} A=-2+\frac{1}{d}
%\ee
%The solution $A=0$ requires $D$ to take imaginary values, either for $D$ or $H_0$. For consistency we choose $A=-2+1/d$. Once $D$ is determined we obtain the solution
\begin{eqnarray}
\varphi &=& \Big(-2+\frac{1}{d}\Big)\log t + B \label{Solution-onlyH-phi} \\
\lambda &=& \frac{1}{d}\log t +\frac{1}{2d}\log \Big(\frac{H_0^2 d^2}{d-1}\Big) \label{Solution-onlyH-lambda}
\end{eqnarray}
This analytic solution is plotted as gray colored straight lines in figure \ref{lambda-late-h} for different values of $H_0$. Notice that $H_0$ fixes the initital value of  $\lambda (t=1)$ in these solutions.  
%This, as we can notice, determines the moment in which the scale factor begins to  grow in .

%%%%%%%%%%%%%%%%%
%%%%%%%%%%%%%%%%%

% In this case, if we considerNevertheless, for the $H_0\neq 0$ the  contrast, if we turn on the two-form field, after some time the term containing the $U(\lambda)$ potential becomes dominant and drives the expansion of the scale factor. Also, it alters the late time behavior of the shifted dilaton $\varphi$, as can be seen on figure \ref{lambda-late-b}, the slope of the Log-linear plot is changed by the presence of the form flux.

%: figure: no winding, no momentum, h

In this case we find that $H_0$ is able, by itself, to induce the growth of a large scale factor, as can be seen in figure \ref{lambda-late-h-a}. In this figure we can see how the two-form field flux induces decompactification for different values of $H_0$.  Since the two-form field flux is along 3 spacial dimensions, in the absence of winding and momentum modes, this field alone is able to induce the growth of 3 large spacial dimensions.  We also notice that the moment in which the scale factor is able to "escape" the constant solution depends on the value of $H_0$.  For larger values of it, the scale factor begins to increase earlier.

This kind of scenario, in which the two-form field flux happens to be the dominant term, can occur if the pressure coming from the winding and momentum modes becomes  negligible ($P_\lambda \approx 0$). This happens in generic situations, for example, when the scale factor remains near the self-dual radius, the dilaton goes to weak coupling or when the winding and  momentum modes have annihilated.

%%%%%%%%%%%%%
%%%%%%%%%%%%%%%%%
%%%%%%%%%%%%%%%%%%%%%
\subsubsection{$H_0\neq 0$, $w\neq 0$ case}
%%%%%%%%%%%%%%%%%%%%%
%%%%%%%%%%%%%%%%%
%%%%%%%%%%%%%
Finally we investigate the generic case when both the flux and the matter pressure are present.
In order to find a solution when the antisymmetric tensor potential is present and the pressure fulfills the equation of state $P_\lambda=wE$, we use  (\ref{w-eqmotphi}), (\ref{w-eqmotlambda})  and the ansatz (\ref{latetimecond}). 
Keeping only up to quadratic terms, we find
\bea
-\frac{A}{t^2} &=&\frac{A^2}{2t^2}+\frac{dC^2}{2t^2}+\frac{H_0^2}{2}e^{-2d(C\log t +D)}~\\
-\frac{C}{t^2} &=& \frac{AC}{t^2}+H_0^2 e^{-2d(C\log t +D)}+\frac{w}{2}\left(\frac{A^2}{t^2}-\frac{dC^2}{t^2}-H_0^2 e^{-2d(C\log t +D)}\right)~
\eea
By supposing that $C=\frac{1}{d}$, we can eliminate  the $t^{-2}$ dependence on the equations. We obtain then
\bea
-A &=&\frac{A^2}{2}+\frac{1}{2d} +\frac{H_0^2}{2}e^{-2dD} \\
-\frac{1}{d}&=&\frac{A}{d}+H_0^2e^{-2dD}+\frac{w}{2}(A^2-\frac{1}{d}-H_0^2e^{-2dD}) 
\eea
Substituting $d=3$ explicitly and solving for $A$ and $D$, we find
\bea
A&=&\frac{5- 3w}{3(w-1)} \\
H_0^2 e^{-6 D}&=&\frac{2-12w+6w^2}{9(w-1)^2} \label{H_o-positive}
\eea
In this way we find a solution
\bea
\varphi &=&\left(\frac{5- 3w}{3(w-1)}\right)\log t + B \label{phi-rad-h}\\
\lambda &=& \frac{1}{3}\log t -\frac{1}{6}\log \left( \frac{2-12w+6w^2}{9H_o^2(w-1)^2}\right)~ \label{lambda-rad-h}
\eea
We observe, on equation (\ref{H_o-positive}) that $w$ is constrained by the inequality
\be
w<1-\sqrt{\frac{2}{3}}\approx 0.1835~
\ee
It is not consistent with $w=1/3$ (radiation).
This problem indicates that we cannot smoothly connect this solutions to those with $H_0=0$.

%%%%%%%%%%
%%%%%%%%%%
\subsection{Perturbative solutions}\label{sec: perturbative}

\subsubsection{$H_0\neq 0$, $w\neq 0$ case}\label{sec:effectearly}
Due to the difficulty we just encountered, we construct perturbative solutions with non-vanishing flux starting from those with
no flux.
Using the solutions we have obtained for the case when $H_0=0$, we treat  the potential term due to $H_0\neq 0$ as a perturbation to the equations of motion. The small expansion parameter is
\be
\delta\equiv H_0^2 
\ee
We expand the solution in terms of the small parameter $\delta$
\bea
\varphi &=&\varphi_0 +\delta\varphi_1 +\delta^2\varphi_2+\cdots \label{phi-delta}\\
\lambda&=&\lambda_0+\delta\lambda_1+\delta^2\lambda_2+\cdots\label{lambda-delta} \eea 
and substitute (\ref{phi-delta}), (\ref{lambda-delta}) into the equations of motion.
They describe perturbations around the solutions $\varphi_0$, $\lambda_0$ obtained in (\ref{phi-latetime-no-h}), (\ref{lambda-latetime-no-h}).
%\bea
%\varphi_0 &=&-\frac{2}{1+d w^2}\log t +B~ \\
%\lambda_0 &=&+\frac{2w}{1+dw^2}\log t +D~
%\eea

From the power series expansion of the equation of motion (\ref{lambdaeqmot}), we get the differential equation for the first order terms in $\delta$
\bea
\ddot\varphi_1 &=& \dot{\varphi_0}\dot\varphi_1 +d\dot{\lambda_0}\dot\lambda_1+\frac{1}{2}e^{-2d\lambda_0}   \\
\ddot\lambda_1 &=& \dot{\varphi_0}\dot\lambda_1 + \dot{\lambda_0}\dot\varphi_1+w(\dot{\varphi_0}\dot\varphi_1-\dot{\lambda_0} \dot\lambda_1 )+d e^{-2d\lambda_0}(1-\frac{w}{2})~
\eea
After substituting $\varphi_0$, $\lambda_0$, $\dot{\varphi_0}$, $\dot{\lambda_0}$ into the equation, we obtain
\bea
\ddot\varphi_1 &=& -\frac{2}{1+dw^2} t^{-1}\dot\varphi_1  +\frac{2d w}{1+dw^2} t^{-1}\dot\lambda_1 +\frac{1}{2} t^{-\frac{4dw}{1+d w^2}} e^{-2dD} ~\\
\ddot\lambda_1 &=& -\frac{2}{1+dw^2}t^{-1}\dot\lambda_1 +\frac{2w}{1+dw^2}t^{-1}\dot\varphi_1 +w(-\frac{2}{1+dw^2}t^{-1}\dot\varphi_1 - \frac{2dw}{1+d w^2}t^{-1}\dot\lambda_1) \nonumber\\ && \quad\quad\quad + d t^{-\frac{4dw}{1+dw^2}}e^{-2dD}(1-\frac{w}{2})~
\eea
For $\lambda_1$,  we get a second order differential equation in this way
\be
\ddot\lambda_1 +2t^{-1}\dot\lambda_1 - d  t^{-\frac{4dw}{1+dw^2}}e^{-2dD}(1-\frac{w}{2})=0~
\ee
We can integrate this equation easily. Defining $x\equiv\dot\lambda_1$, $\dot x\equiv\ddot\lambda_1$ we get
\be
\dot x+2t^{-1}x=t^{-\frac{4dw}{1+dw^2}}e^{-2dD}d(1-\frac{w}{2})~
\ee
This is a differential equation of the form $\dot x(t)+f(t)x(t)=g(t)$ and the solution is given by\be
x(t)=\frac{\int dt ~g(t)e^{\int f(t)dt}+c}{e^{\int f(t)dt}}~
\ee
with a constant $c$. 
%Then,
%\be
%x=\frac{\int dt~( t^{-\frac{4dw}{1+dw^2}}e^{-2dD}d)(1-\frac{w}{2})t^2+c}{t^2}~.
%\ee

After the integration, we find two different class of solutions: 
\begin{itemize}
\item $\frac{4dw}{1+dw^2}\neq 3$ case.
\be
x=d e^{-2dD}(1-\frac{w}{2})(3-\frac{4dw}{1+dw^2})^{-1}t^{-\frac{4dw}{1+dw^2}+1}+c t^{-2}
\ee
\be
\lambda_1 = d e^{-2dD}(1-\frac{w}{2})(3-\frac{4dw}{1+dw^2})^{-1}(2-\frac{4dw}{1+dw^2})^{-1}t^{-\frac{4dw}{1+dw^2}+2}-ct^{-1}+c_0
\ee
Here, the leading perturbation contains two different time dependent terms. For the perturbation to be small, the exponent on the first term should fulfill the condition 
\be
-\frac{4dw}{1+dw^2}+2 <0~
\ee
%This is the case for $d>3$ and a universe filled by radiation ($w=1/d$). 
If it is the case, the influence of the two-form flux induced potential is negligible in comparison to the pressure of the string momentum modes. 
On the other hand, this condition is not satisfied for $w=0$ (pressureless dust) case where the perturbation grows as $t^2$. In such a situation the solution $\lambda_0$ is unstable and the universe is decompactified due to the presence of the two-form field flux.
\item $\frac{4dw}{1+dw^2}= 3$.
This is the case for $d=3$ and a universe filled with radiation ($w=1/d$).
\be
x=s\frac{\log t}{t^2}+\frac{c}{t^2}
\ee   
\be
\delta\lambda = -\frac{s+c}{t}-s\frac{\log t}{t}+c_0~, \qquad s=-\frac{5}{2}H_0^2 e^{-6D}~
\ee
When $t\rightarrow\infty$, we find the leading perturbation as $\delta\lambda \sim\mathcal{O}(t^{-1})$ . Therefore the correction to the unperturbed solution is negligible at late time.
\end{itemize}

%Substituting this solution in the expression for  $\delta\ddot\varphi$
%\be
%\delta\ddot\varphi =-\frac{2}{1+d w^2}t^{-1}\delta\dot\varphi +\frac{2dw}{1+dw^2}t^{-1}[ h(2-\frac{4dw}{1+dw^2})t^{-\frac{4dw}{1+dw^2}+ct^{-2}}
%\ee

%%%%%%%%%%%%%%%%%%%%
%%%%%%%%%%%%%%%%%%%%
\subsubsection{$H_0\neq 0$ case with both momentum and winding modes}\label{sec:effectearly}
As we observe in equation (\ref{eqmotlambda}), the pressure coming from the winding modes and the momentum modes is multiplied by $e^\varphi$. If $|H_0|\ll 1$  and $|\varphi |\approx 1$,  the scale factor experiences oscillations in the presence of winding and momentum modes. As $\varphi$ goes to weak coupling, oscillations stop and the  pressure terms become small with respect to the $H_0^2$ potential term. Before  this terms becomes significant, the solution is characterized as
\be
\dot\varphi\approx 0,~\dot\lambda\approx 0 ,~\lambda \approx 0, e^\varphi\ll 1
\ee
We define a small parameter
\be
\epsilon \equiv \frac{H_0^2}{2^d}
\ee
Under this approximation, keeping terms to the lowest nontrivial order, we get
\bea
\ddot\varphi &=&\frac{1}{2}(\dot\varphi^2+d\dot\lambda^2+\epsilon) \\
\ddot\lambda &=&\dot\varphi\dot\lambda +\epsilon\lambda
\eea
We expand the solution in terms of the small parameter $\epsilon$
\bea
\varphi &=&\varphi_0 +\epsilon\varphi_1 +\epsilon^2\varphi_2+\cdots \label{phi-serie}\\
\lambda&=&\lambda_0+\epsilon\lambda_1+\epsilon^2\lambda_2+\cdots\label{lambda-serie} \eea 
After substituting (\ref{phi-serie}), (\ref{lambda-serie}) into the equation of motion, 
%\bea
%\ddot\varphi_0 +\epsilon\ddot\varphi_1 &=&\frac{1}{2}(\dot\varphi_0^2 + 2\epsilon\dot\varphi_0\dot\varphi_1)+\frac{1}{2} d(\dot\lambda_0^2 +2\epsilon\dot\lambda_0\dot\lambda_1)+\frac{\epsilon}{2}\\
%\ddot\lambda_0 +\epsilon\ddot\lambda_0 &=&\dot\varphi_0\dot\lambda_0 +\epsilon (\dot\varphi_0\dot\lambda_1 +\dot\lambda_0\dot\varphi_1)+\epsilon (\lambda_0 +\epsilon\lambda_1)
%\eea
we have  differential equations at each order of the perturbation parameter
\bea
\ddot\varphi_0 &=&\frac{1}{2}\dot\varphi_0^2 +\frac{1}{2}d\dot\lambda_0^2 \\
\ddot\lambda_0 &=&\dot\varphi_0\dot\lambda_0 \\
\ddot\varphi_1&=&\dot\varphi_0\dot\varphi_1 +d\dot\lambda_0\dot\lambda_1+\frac{1}{2} \label{perturbationphi}\\
\ddot\lambda_1 &=& \dot\varphi_0\dot\lambda_1 +\dot\lambda_0\dot\varphi_1 +\lambda_0 \label{perturbationlambda}
\eea
where the solutions for $\varphi_0$, $\lambda_0$ is given in equation (\ref{phi-latetime-no-h}), (\ref{lambda-latetime-no-h}). 
Substituting them in (\ref{perturbationphi}), (\ref{perturbationlambda}), we obtain
\bea
\ddot\varphi_1&=&-\frac{2}{t}\dot\varphi_1 +\frac{1}{2} \\
\ddot\lambda_1 &=& -\frac{2}{t}\dot\lambda_1 +c_0
\eea
with $c_0$ a constant. These equations are linear differential equations in $\dot\varphi_1$ and $\dot\lambda_1$ respectively. They can be solved by multiplying them by the integrating factor $e^{(\int dt\frac{2}{t})}=e^{(2\log t)}=t^2$.
The solutions are
\bea
\dot\lambda_1(t)&=&\frac{\lambda_0}{3}t +(const.)  \quad\quad \lambda_1(t)=\frac{\lambda_0}{6}t^2 +(const.)t^{-1} + const. \\
\dot\varphi_1(t)&=&\frac{1}{6}t+(const.)\quad\quad    \varphi_1(t)=\frac{1}{12}t^2 +(const.)t^{-1}+const.
\eea

\begin{figure}
\begin{center}
\subfigure[]{
\includegraphics[width=7cm,height=6cm]{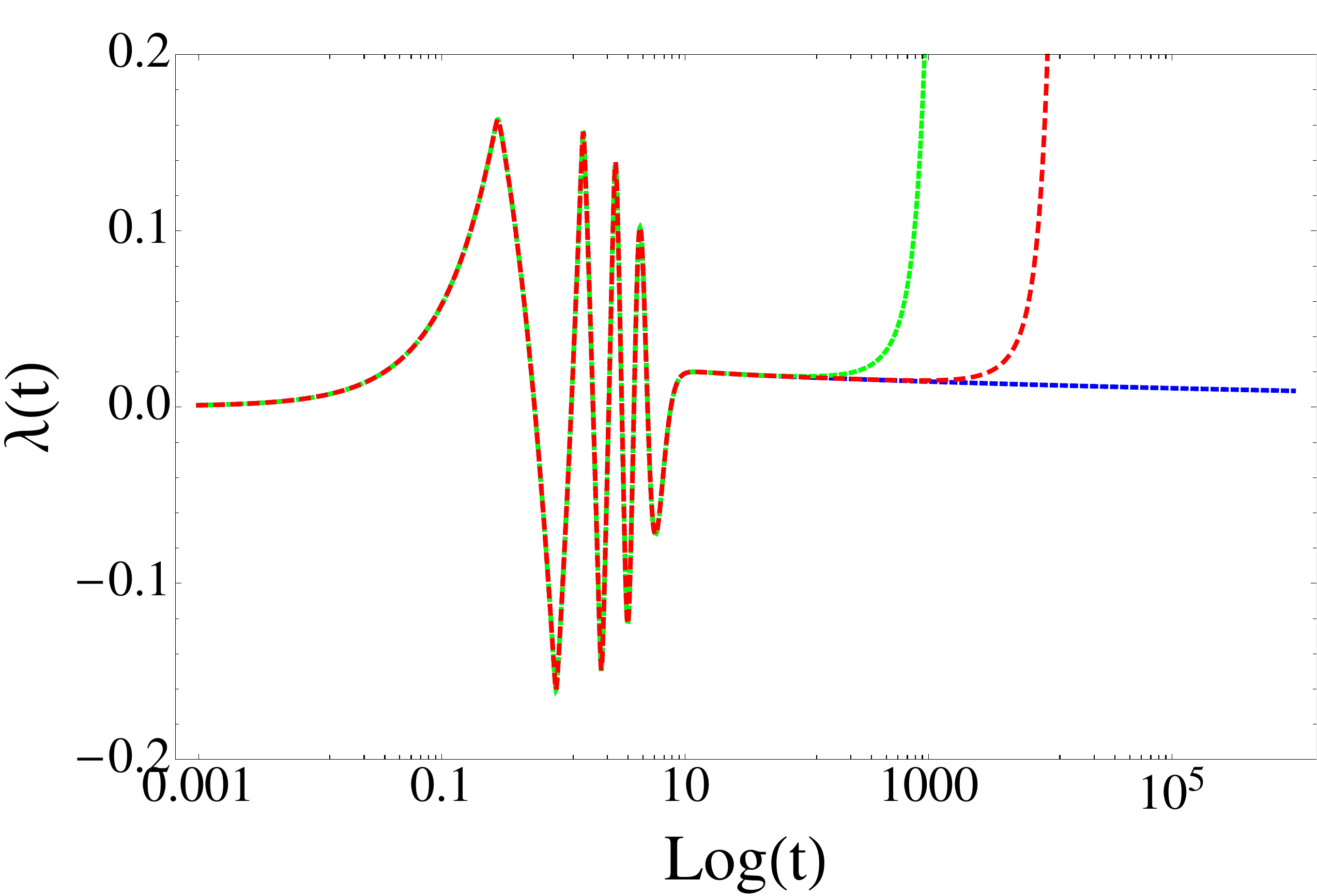}}~~~~~~~
\subfigure[]{\includegraphics[width=7cm,height=6cm]{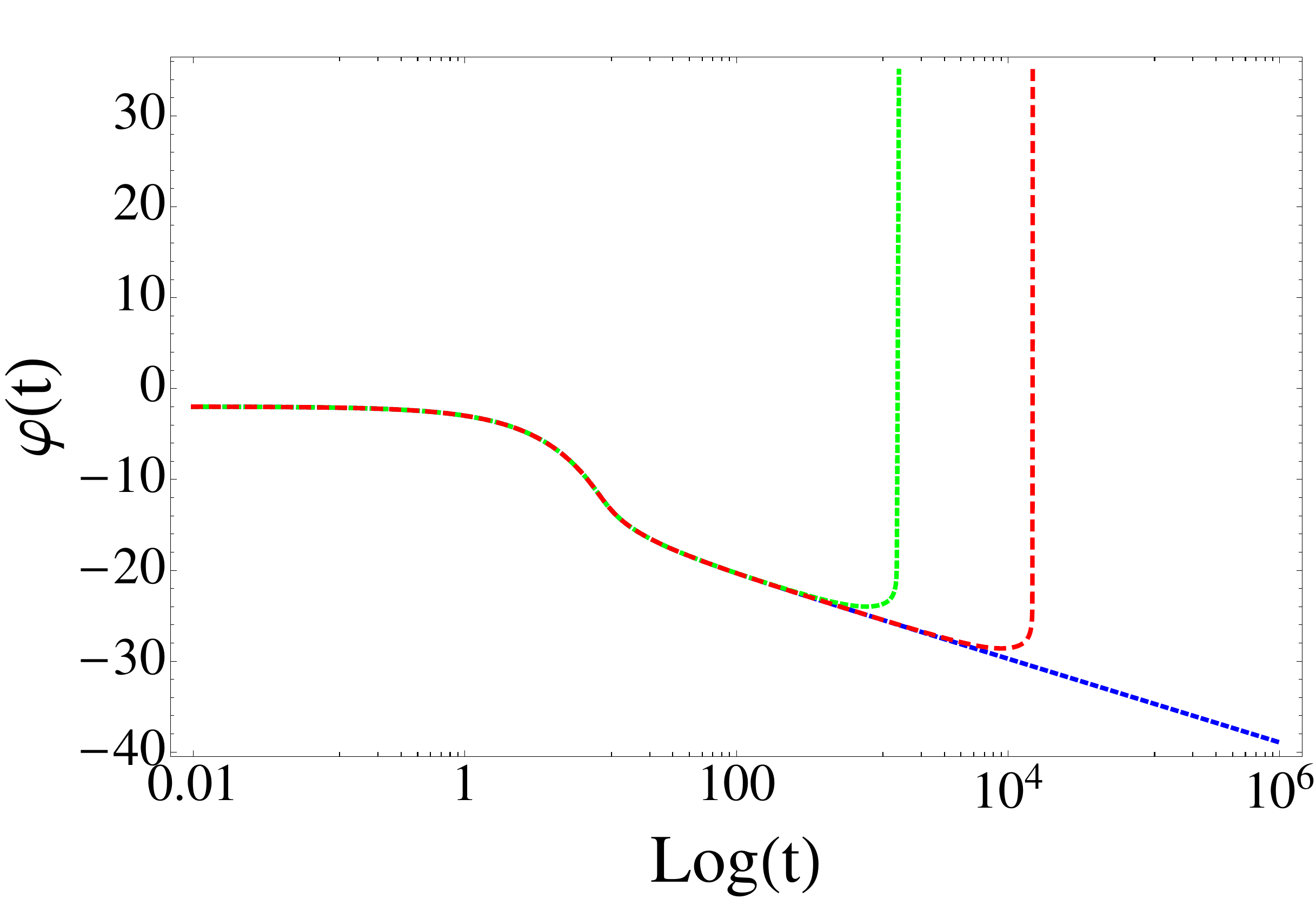}}
 \end{center}
\caption{behavior of $\lambda$ in presence of $U(\lambda)$, winding and momentum modes ($W\neq 0, K\neq 0$, {\color{NavyBlue}$H_0=0$}, {\color{Green}$H_0=0.01$}, {\color{Red}$H_0=0.001$} ).}\label{momentum+winding+H}
\end{figure}

We observe the following features: the solution $\varphi_0,~ \lambda_0$ becomes unstable if we perturbed it with nonvanishing $H_0$.
To leading order the solution in this regime behaves like $\sim t^2$. 
This instability initiates an accelerated expansion of a universe away from the oscillating phase around the self-dual radius.
However
we also observe that the perturbation also affects the dilaton. As the perturbation becomes dominant, the dilaton begins to grow and goes to strong coupling.  This indicates that a bounce on the dilaton has been produced. A result like this looks problematic, since a bounce on the dilaton leads to a violation on the positive energy condition as was noted in \cite{Greene:2008hf}. We may not be able to trust our solution there as it also takes the dilaton to strong coupling. This behavior can be observed directly in a numerical solution of the equations (figure \ref{momentum+winding+H}). We begin with a string gas of equal number of winding modes and momentum modes. Before the dilaton goes to weak coupling, the scale factor oscillates around the self-dual radius. Once the dilaton reaches weak coupling region, the oscillations stop and the scale factor stabilizes. Then the potential induced by the two-form field flux becomes dominant and the solution begins to grow as predicted by the perturbed solution.
In the next sub-section, we investigate the effects of the string interaction on these problems through the Boltzmann equations.

 %the dilaton will begin to increase towards strong coupling and, depending on the value of $\lambda_0$, te scale factor will begin to increase or decrease, depending on the value of $\lambda_0$ when the $H_0$ term becomes dominat. This behavior can be seen in figure \ref{momentum+winding+H}. Both solutions are equivalent, because of T-duality. This phenomenon is easy to understand since the potential we are introducing  has a local maximun at $\lambda =0$ (figure \ref{Vlambda}), then, the $H_0$ term makes the scale factor grow at early times. Also, we can notice that the dilaton begins to go to strong coupling again. 
 
%: figure: Winding and momentum together with H field, no Boltzmann
%%%%%%%%%%%%%%%
%%%%%%%%%%%%%%%%
\subsection{Effects of string interactions}
%(Boltzmann equations)
%%%%%%%%%%%%%%%%%%%
Up to this moment, we have considered situations in which  the winding and momentum numbers are frozen at their initial values.  When the string gas falls out of equilibrium in an expanding universe, winding strings in the gas can interact and begin to annihilate. In this section we incorporate, together with the two-form field flux induced potential, the Boltzmann equations that take account of the interaction among strings. These equations, derived by Polchinski \cite{Polchinski:1988cn}, are shown below %in let's consider the  coupling of the equations of motion for interacting matter.  Now, we present  the equations of motion for the interaction of strings at weak coupling \cite{Polchinski:1988cn}
%: figure: Boltzmann with momentum and winding
\begin{figure}[htp]
\begin{center}
\subfigure[$\lambda (t)$] {\label{Boltzmann-lambda}
\includegraphics[width=7cm,height=6cm]{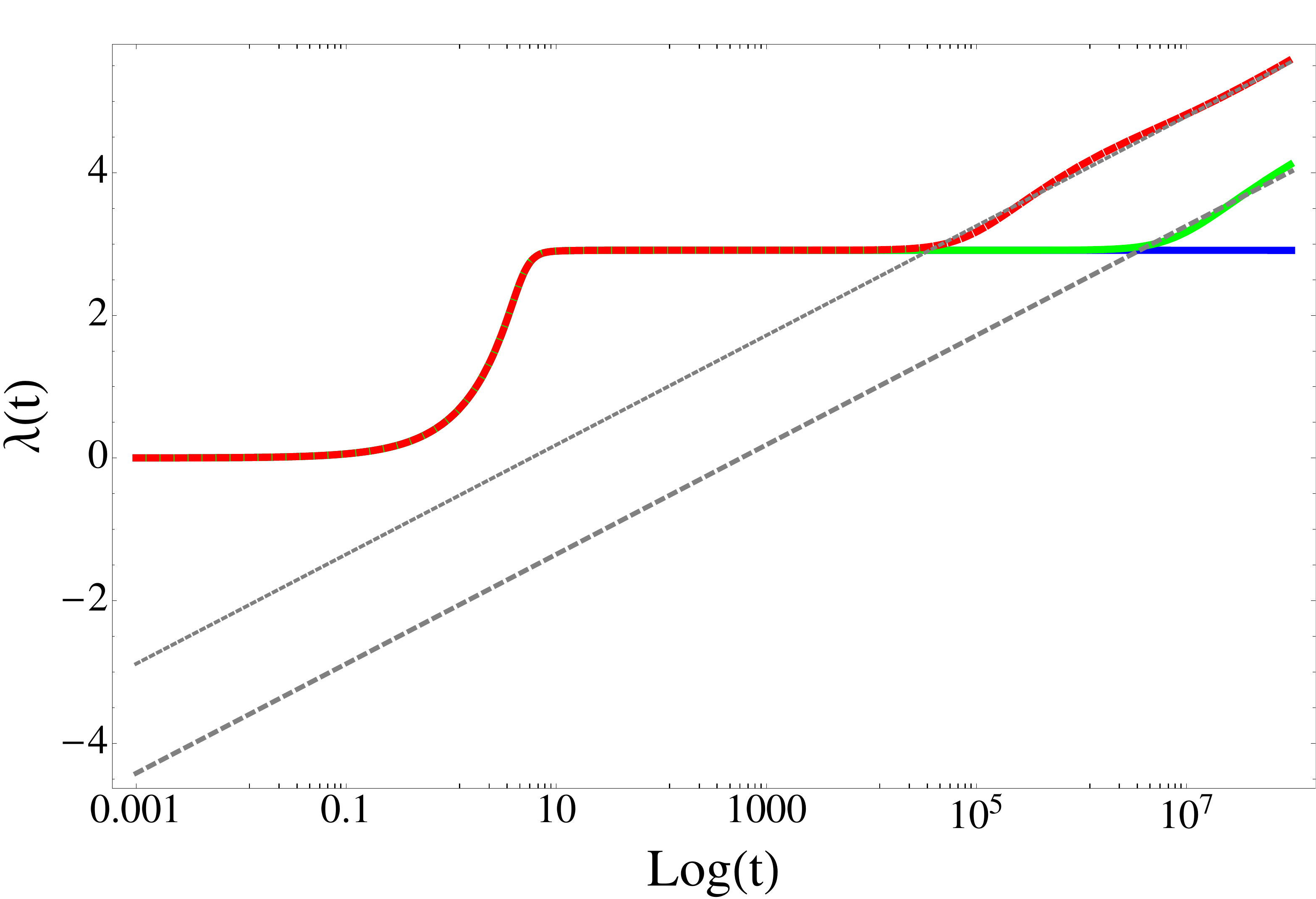}}\quad
\subfigure[$\varphi (t)$] {\label{Boltzmann-phi}
\includegraphics[width=7cm,height=6cm]{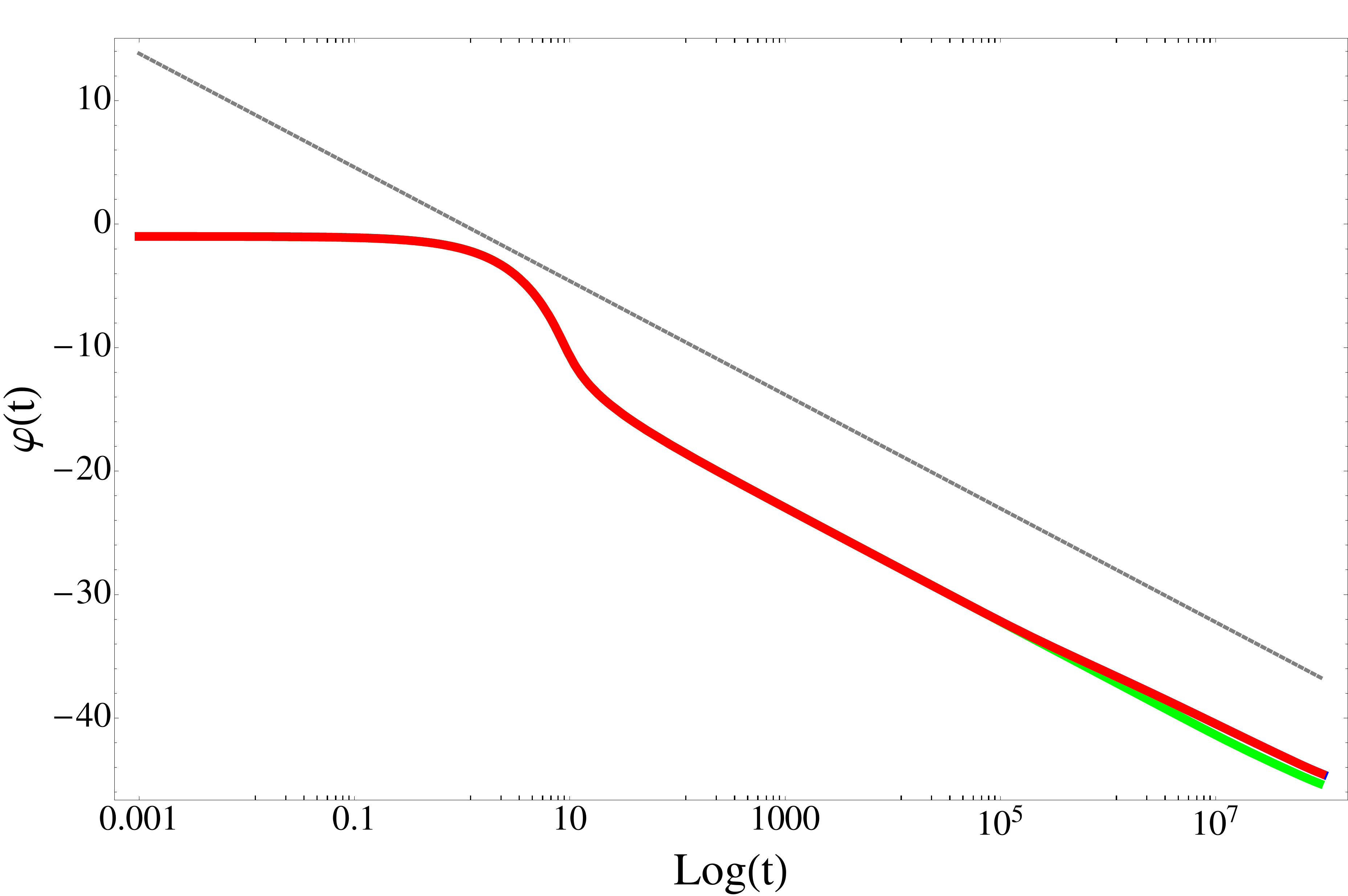}}\quad
\subfigure[$K(t),~W(t)$] {\label{Boltzmann-k-w}
\includegraphics[width=7cm,height=6cm]{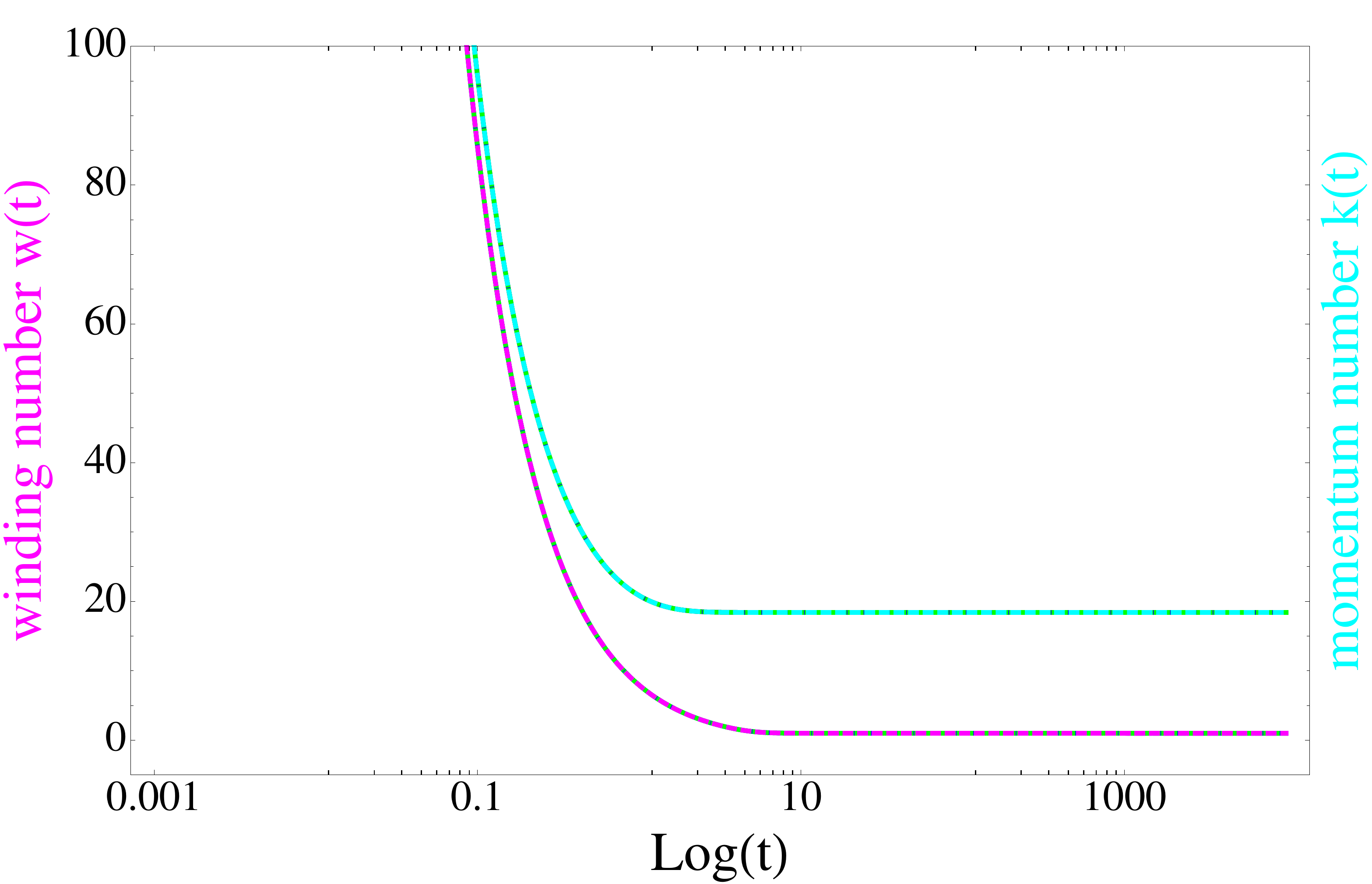}}
\end{center}
\caption{(a), (b) behavior of $\lambda$ and $\varphi$ with initial condition $K=W$ for  {\color{NavyBlue} $H_0=0$}, {\color{Green}$H_0=0.001$} and  {\color{Red}$H_0=0.1$}. The  effect of the Boltzmann equations is included. (c) Evolution of the winding number and momentum number is plotted (the solutions overlap for the three cases considered).}
\label{Boltzmann-k-w-lambda}.
\end{figure}
\bea
\dot W &=&-\frac{e^{2\lambda +\varphi}}{\pi}(W^2-\langle W\rangle^2)~\label{BoltzmannW}\\
\dot K&=&-\frac{e^{-2\lambda +\varphi}}{\pi}(K^2-\langle K\rangle^2)~\label{BoltzmannK}
\eea
We combine these equations with (\ref{eqdilaton}), (\ref{eqmotlambda}) and evolve the system numerically. The universe  we consider is filled with gas of strings that begins at the self dual radius with equal initial winding and momentum numbers ($K=W$). The initial conditions are $\dot\lambda\approx 1$, $\dot\varphi\approx -1$ and the dilaton is going from strong coupling to weak coupling. The numerical results including the effects of the Boltzmann equations are presented in  figure \ref{Boltzmann-k-w-lambda}. We have plotted the behavior of the scale factor $\lambda(t)$, the dilaton $\varphi(t)$, the winding number $w(t)$ and the momentum number $k(t)$. In figure \ref{Boltzmann-k-w-lambda} we  observe that, as $\lambda(t)$ grows, the winding modes begin to annihilate. Then, there is not enough pressure to make the universe contract and experience bounces. Instead, the contribution from the two-form field becomes dominant and the scale factor tends to the solution  (\ref{Solution-onlyH-phi}), (\ref{Solution-onlyH-lambda}) with vanishing pressure where the scale factor grows large due to the flux induced potential.

The behavior of the winding and the momentum number is as expected from the following characteristics of the Boltzmann equations (\ref{BoltzmannW}), (\ref{BoltzmannK}). As the dilaton goes to weak coupling, the interaction rate goes to zero and the values of the winding and momentum numbers become constant.  When the scale factor grows large, winding modes annihilate more efficiently because their interaction rate goes as the exponential of the scale factor. On the contrary, the rate of annihilation  of the momentum modes  becomes smaller because the interaction rate between them decays exponentially with the scale factor. In fact this asymmetry between  winding and momentum modes can be observed in figure \ref{Boltzmann-k-w}. 

The result obtained by taking account of the Boltzmann equations suggests an interesting scenario when homogeneous $H_{\mu\nu\lambda}$ is present. If the winding modes annihilate rapidly enough, the effect of 2-from field flux becomes important even at early times. The annihilation of the winding modes could take place even in a loitering phase. In that case the two-form field flux becomes dominant and the  the expansion of three large spacial dimensions is realized.
We emphasize that this mechanism is different from Brandenberger-Vafa mechanism as the presence of homogeneous 2-form field flux is crucial for three spacial dimensions to grow. Without it, the universe remains to be of microscopic size as the blue line in \ref{Boltzmann-lambda} indicates.

%%%%%%%%%%%%%%%%%%
%%%%%%%%%%%%%%%%%%
%%%%%%%%%%%%%%%%%%

\section{Effects of constant $B_{\mu\nu}$}\label{sec:constant_field}
\setcounter{equation}{0}
So far, the effect of the two-form field has entered only as a modification to the usual dilaton gravity action, as in \cite{Campos:2003ip}. The string gas model, as it stands, couples the modified action of dilaton gravity with that of a gas of strings. In this approach the effect of the background field $B_{\mu\nu}$ over the string spectrum is usually neglected. The correction on the energy of the string goes as $\mathcal{O} (B)$, thus, this approximation is valid for weak fields.
In dilaton-gravity,  the contribution of $B_{\mu\nu}$ to the action enters via $U(\lambda)\sim |H|^2=|dB|^2$. In this case, even if $B$ remains small, $H$ is not necessary so as the space-time variation of $B_{\mu\nu}$ could be large.  
%Moreover, the action \ref{} the contribution of $B_{\mu\nu}$ enters as a potential $U(\lambda)\sim |H|^2=|dB|^2$, such that the spatial variation of  the two-form field should be constrained to be small. 

In principle, if we know the two-form field in terms of the scale factors, we can determine $|H|^2$ as well as their effect on the string spectrum.  We can then make use of the adiabatic approximation to study the time dependence of the compactification radii and get the equations of motion.  In practice, a homogeneous solution for supergravity is given in terms of $H$. This presents a  problem since we need $B_{\mu\nu}$, not $H$, in order to get the string spectrum.

With this prospect, we investigate the simplest case, that of a constant $B_{\mu\nu}$. In this case, the $H$ dependent term on the supergravity action vanishes as well as the contribution to the equations of motion. Nevertheless, since strings carry charge under the gauge field, the effect of $B$ field on closed strings wrapping the compact dimensions is non-trivial.
%The question is how small it should be?  I don't know, this is rubish, argh...

The Polyakov action in the presence of an antisymmetric field $B_{\mu\nu}$
\be
S=-\frac{1}{\pi\alpha^\prime}\int d^2\sigma[\partial_a X^\mu\partial^a X_\mu -\epsilon^{ab}B_{\mu\nu}\partial_\alpha X^\mu\partial_b X^\mu]
\ee
yields the equations of motion
\be
(\partial_\tau^2-\partial_\sigma^2)X^\mu(\sigma,\tau)=-\frac{1}{2}{H^\mu}_{\lambda\nu}\epsilon^{ab}\partial_a X^\lambda\partial_b X^\nu
\ee
with 
$
H_{\mu\nu\lambda}\equiv \partial_\mu B_{\nu\rho}+\partial_\rho B_{\mu\nu}+\partial_\nu B_{\rho\mu}~.
$
Then, for a constant $B_{\mu\nu}$ we obtain the usual two dimensional wave equation
\bea
(\partial^2_\tau-\partial^2_\sigma)X^\mu=0~
\eea
that allows us to give the solution in a Fourier-Laurent expansion 
\be
\dot X^\mu =\sqrt{\frac{\alpha^\prime}{2}}(\tilde\alpha_0+\alpha_0)+\sqrt{\frac{\alpha^\prime}{2}}\sum_{n\neq 0}(\tilde\alpha_n e^{-in(\tau +\sigma)} +\alpha_n^\mu e^{-in(\tau-\sigma)})\label{Xdot}
\ee
\be
X^{\prime\mu}=\sqrt{\frac{\alpha^\prime}{2}}(\tilde\alpha_0-\alpha_0)+\sqrt{\frac{\alpha^\prime}{2}}\sum_{n\neq 0}(\tilde\alpha_n e^{-in(\tau +\sigma)} -\alpha_n^\mu e^{-in(\tau-\sigma)})~ \label{Xprime}
\ee
%From (\ref{Xdot}) and (\ref{Xprime}) we can determine the worldsheet energy-momentum tensor. 
As it turns out the zero-modes are the only ones that are affected by the  $B_{\mu\nu}$ field. The components of the energy momentum tensor and their zero modes are given by
\bea
T^{01}&=&\frac{1}{2\pi\alpha^\prime}\partial^0 X_\mu\partial^1 X^\mu\nonumber\\
&=&\frac{1}{2\pi\alpha^\prime}\bigg[ \alpha^\prime p^\mu - {B^\mu}_j(wR)^j +\sqrt{\frac{\alpha^\prime}{2}}\sum_{n\neq 0}(\tilde\alpha_n^\mu e^{-in(\tau +\sigma)}+\alpha_n^\mu e^{-in(\tau-\sigma)}) \bigg]\nonumber\\
&&\times\bigg[(wR)_\mu+\sqrt{\frac{\alpha^\prime}{2}}\sum_{n\neq 0}(\tilde\alpha_{n\mu} e^{-in(\tau +\sigma)}-\alpha_{n\mu} e^{-in(\tau-\sigma)})\bigg] ~
 \eea
 
 \be
 (T^{01})_{\text{zero~modes}}=n^iw_i +\frac{1}{2}\sum_{n=1}(\tilde\alpha_n\cdot\tilde\alpha_{-n}+\tilde\alpha_{-n}\cdot\tilde\alpha_n-\alpha_{-n}\cdot\alpha_n-\alpha_n\cdot\alpha_{-n})
 \ee
 
 \bea
T^{00}&=&\frac{1}{4\pi\alpha^\prime}(\dot X\cdot\dot X+X^\prime\cdot X^\prime) \nonumber\\
&=&\frac{1}{4\pi\alpha^\prime}\bigg[\bigg(\alpha^\prime p^\mu - {B^\mu}_j(wR)^j +\sqrt{\frac{\alpha^\prime}{2}}\sum_{n\neq 0}(\tilde\alpha_n e^{-in(\tau +\sigma)}+\alpha_n^\mu e^{-in(\tau-\sigma)})\bigg)^2\nonumber\\
&&+\bigg((wR)^\mu+\sqrt{\frac{\alpha^\prime}{2}}\sum_{n\neq 0}(\tilde\alpha_n^\mu e^{-in(\tau+\sigma)}-\alpha_n^\mu e^{-in(\tau-\sigma)})\bigg)^2\bigg]
\eea
\bea
(T^{00})_{\text{zero~modes}}&=&\frac{1}{\alpha^\prime}\bigg[-(\alpha^\prime p^0-{B^0}_j(wR)^j)^2+(\alpha^\prime p^i -{B^i}_j(wR)^j)^2+(wR)^i(wR_i)\nonumber\\
&&+\alpha^\prime\sum_{n=1}(\tilde\alpha_n\cdot\tilde\alpha_{-n}+\tilde\alpha_{-n}\cdot\tilde\alpha_n+\alpha_n\cdot_{-n}+\alpha_{-n}\cdot\alpha_n)\bigg]
\eea
Imposing the physical constraint that the energy momentum tensor must vanish, we get the energy spectrum for the string
%\be
%(\alpha^\prime p^0 -{B^0}_jw^jR)^2 = \alpha^{\prime 2}\Big(\frac{n}{R}\Big)^i\Big(\frac{n}{R}\Big)_i +(wR)^i(wR)^i + {B^i}_j(wR)^j(wR)^k +\alpha^\prime B_{ik}n^iw^k +\alpha^\prime (\tilde N+N)
%\ee
\be
p^0=\frac{B^0_j w^jR}{\alpha^\prime}+\frac{1}{\alpha^\prime}\sqrt{\alpha^{\prime 2}\Big(\frac{n}{R}\Big)^i\Big(\frac{n}{R}\Big)_i +  (wR)^i(wR)_i +{B^i}_jB_{ik}(wR)^j(wR)^k+\alpha^\prime B_{ik}n^iw^k+\alpha^\prime(\tilde N+N)} \label{spectrum}
\ee
and the level matching condition
\be
\tilde N -N=n^iw_i~~
\ee
%In this equations we have considered any ordering constant to be contained on $N$, $\tilde N$. 
Since all the spacial coordinates are compactified with radius $R^i$, the momentum is quantized as $p^i=(n/R)^i$, where $i$ denotes the spacial index. 

In order to be able to solve the equations of motion, we need  to assume some initial winding and momentum distribution of the string gas. 
%Nevertheless, as a first approximation to this problem we will assume a given initial configuration. 
%There are two interesting cases that we can treat separately: an electric field and a magnetic field as we will see in the next subsection.
The constant $B_{\mu\nu}$ field could be either electric or magnetic type. 
We find that the effect of electric type field is very interesting as there is a critical 
value for which the string tension vanishes for winding modes.

\subsection{Constant electric type field}
Let's consider the case of a homogeneous electric type field in 3-spacial dimensions, with $B\equiv B_{01}=B_{02}=B_{03}$. In order to demonstrate the most dramatic effect, we assume that strings are aligned in the direction of the electric type field. If this is the case, from (\ref{spectrum}), the energy for the winding modes in (\ref{energy-windingmodes}) is modified as
\be
E_W=2d(1-\sqrt{3}B)W e^\lambda~
\ee
with $W$ the winding number and $d=3$. From this equation, we see immediately that  the effect of the field $B_{\mu\nu}$ is to reduce the energy of the winding strings.  Also, it follows that $B$ is constrained to take values
\be
0\leq B\leq \frac{1}{\sqrt{3}}~
\ee
In particular, when the inequality is saturated $B=1/\sqrt{3}$, the energy of the winding modes vanishes. As the pressure they exert also vanishes, the spacial dimensions are expected to expand freely because  of the presence of the momentum modes.
\begin{figure}
\begin{center}
\subfigure[Scale factor $\lambda$]
{\includegraphics[width=6cm,height=5cm]{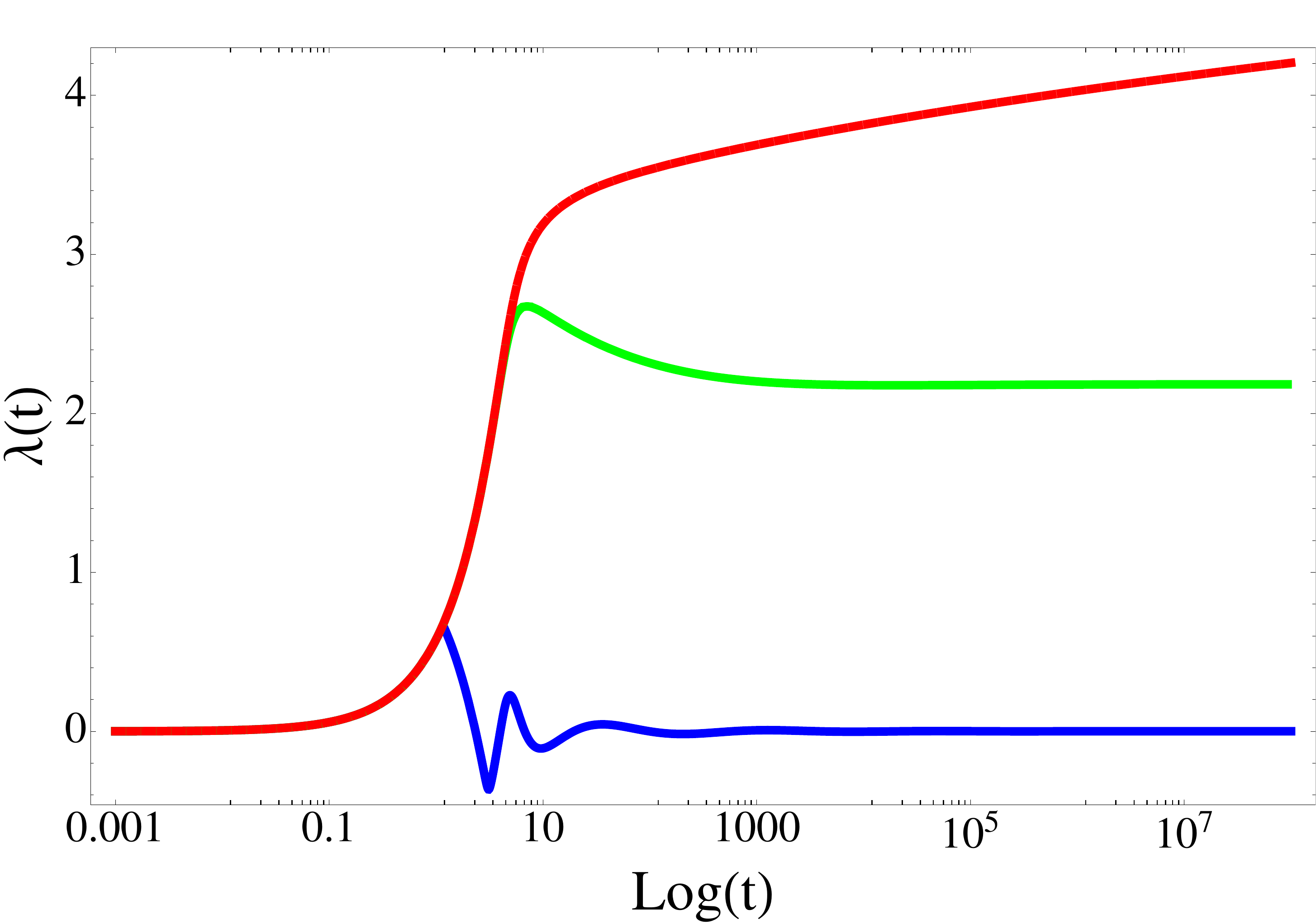}}~~~~~~~
\subfigure[Dilaton $\varphi$] 
{\includegraphics[width=6cm,height=5cm]{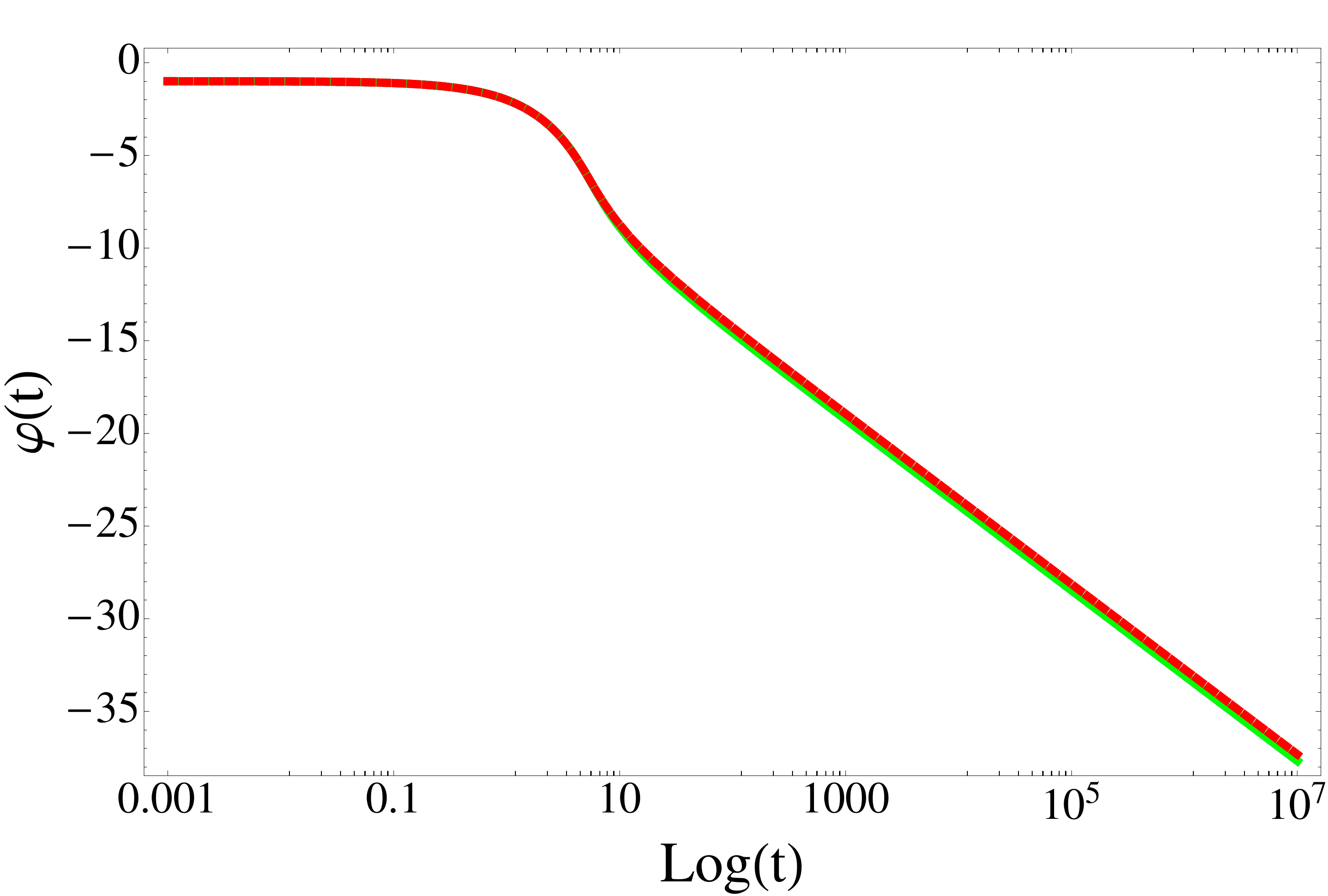}}~~~~~~~
 .
\end{center}\caption{behavior of the dilaton and scale factor for {\color{NavyBlue}$B=0$}, {\color{Green}$B=0.5$}, {\color{Red}$B=1/\sqrt{3}$} (critical electric field) and $W=K$.}\label{fig:electric}

\end{figure}

In figure \ref{fig:electric} we have plotted the numerical solution for different values of $B$ without including the effect of the Boltzmann equations  (\ref{BoltzmannW}). For vanishing $B$ the momentum and winding modes make the scale factor oscillate around the self-dual radius. With small $B\neq0$, the solutions oscillate around positive values of $\lambda$. As we get closer to the critical $B$, the solution bounces and then stabilizes. When we reach the critical value $B=1/\sqrt{3}$, the pressure from the winding modes becomes zero and $\lambda$ expands just like  a universe filled with radiation (momentum modes).

%\begin{figure}
%\begin{center}
%\subfigure[Scale factor $\lambda$]
%{\includegraphics[width=6cm,height=5cm]{lambdaconstelectricfield.pdf}}~~~~~~~
%\includegraphics[width=5cm,height=4cm]{lambdaprimeconstelectricfield.pdf}
%\subfigure[Dilaton $\varphi$] 
%{\includegraphics[width=6cm,height=5cm]{phiconstelectricfield.pdf}}~~~~~~~
%\includegraphics[width=5cm,height=4cm]{lambdaprimeconstelectricfield.pdf}
% \label{fig:electric}.
%\end{center}\caption{behavior of the dilaton and scale factor for critical electric field and vanishing momentum number.}
%\end{figure}

%\subsection{Constant magnetic field}
%For a constant magnetic field we will a configuration given by the the following matrix
%\be
%B_{ij}=
%\left( \begin{array}{ccc}
%0 & B & -B \\
%-B & 0 & B \\
%B & -B & 0 \end{array} \right)
%\ee
%Inserting this into the energy formula in the presence of a magnetic field we get
%\be
%p^0=\frac{1}{\alpha^\prime}\sqrt{B^2[(w_2e^{\lambda_2}-w_3e^{\lambda_3})^2+(w_3e^{\lambda_3}-w_1e^{\lambda_1})^2+(w_3e^{\lambda_3}-w_1e^{\lambda_1})^2]+(w_1e^{\lambda_1})^2+(w_2e^{\lambda_2})^2+(w_3e^{\lambda_3})^2}~.
%\ee
%Consequently, the contribution from the magnetic field becomes null when the winding is homogeneous in all directions. The equilibrium configuration is that with homogeneous windings, that being the case, the effect of this kind of field over the string modes will vanish, but, if  thermal fluctuations generate inhomogeneities in the winding we could expect some effect.

\section{Conclusions}
\setcounter{equation}{0}
In this work we have investigated some effects of the introduction of a two-form field into  the model proposed in \cite{Greene:2008hf}. This  model provides a bouncing and cycling cosmology and also the possibility of long loitering phases. It avoids singularities at finite times but fails to realize three large spacial dimensions from Brandenberger-Vafa mechanism. Having this in mind, we have included a two-form field into the action, since it may provide an alternative mechanism for the decompactification of 3 spacial dimensions. 
We have considered two cases: homogeneous flux $H_{\mu\nu\rho}$ and constant gauge field $B_{\mu\nu}$.% which results induced potential entering the equations of motion and,  a constant $B_{\mu\nu}$, for which case the contribution to the dilaton action vanishes but we can calculate easily the effects on the energy of the string gas.

\subsection{Homogeneous $H_{\mu\nu\rho}$}

In order to make the model  compatible with T-duality, as the string gas model requires, we have adopted a phenomenological modification on the potential induced by the two-form field entering the gravity action. The modified potential is non-singular at $\lambda=0$ and  reduces  to the correct one when the scale factor $|\lambda |$ is large. In addition, it provides a repulsive potential that can make the universe expand.  
%Since the action respects T-duality, we have mainly consider the case of an expanding universe $\lambda\rightarrow\infty$, but the results are equally valid for a contracting universe.

In the investigation  of the behavior of the scale factor and the dilaton under the influence of the two-form field flux, we find two different cases:
\begin{itemize}
\item Matter dominance: \\
At early times the  scale factor can experience bounces as it is governed by the presence of winding and momentum modes. In  section \ref{sec:effectearly} we have observed that the effect of the two form field is not significant at this early stage of the universe. If we assume only the presence of the momentum modes, the late time solutions reduce to those already found in dilaton cosmology. If this solution is perturbed by the introduction of the two-form field flux potential, its influence vanishes as $t\rightarrow\infty$. On that account, this kind of solution is stable under the perturbation and the effect of $H_{\mu\nu\rho}$ is negligible as the universe expands.
%%%
%%%
%%%
\item Two-form potential dominance:\\
We have obtained the late time analytic solution for vanishing matter pressure  and non-vanishing  $H_{\mu\nu\rho}$. This solution corresponds  to an expanding universe, where the initiation time of the expansion  is set by the  parameter $H_0$.  This analytical solution  matches the  leading behavior of the  numerical solution for the equations of motion.

%We found in this model that there are 
In generic situations the contribution of the matter pressure becomes negligible and the scale factor becomes constant. This occurs when the dilaton goes to weak coupling, the oscillations on the scale factor stop or the expansion of the universe comes to a halt. Such a possibility is enhanced  if we consider the effect of interactions between strings.  As momentum and winding modes can annihilate, it drives the pressure to vanish. % value and together with the dilaton going to weak coupling the solution is driven to a constant value.
In all of the above cases, the effect of the matter pressure vanishes and the scale factor becomes approximately constant. Introducing  a $H_{\mu\nu\rho}$ flux, we find that the constant scale factor solution eventually becomes unstable and the scale factor begins to grow as 
\be
\lambda\sim t^2~
\ee  This kind of scenario occurs whenever the dilaton goes to weak coupling and  the scale factor settles to a constant value. This behavior is remarkable, since it produces an accelerated expansion analogous to the inflationary universe. %Another effect that we should remark is that the two-form field perturbation for the dilaton also grows as $t^2$, hence, 
However we also need to address the issue that the perturbation to the dilaton also goes as $t^2$. 
Thus the dilaton may eventually bounce and go to strong coupling. 
%Then, from the energy conditions for bounces we get that the energy in this case must be negative. % for the dilaton is very dramatic since it indicates that the energy must be negative.
The string interaction effect through the Boltzmann equation is observed to resolve these problems as in figure \ref{Boltzmann-k-w-lambda}.
\end{itemize}

\subsection{Constant $B_{\mu\nu}$}
We have also investigated the case of a constant $B_{\mu\nu}$ in order to test how its presence affects the action of the string gas. For a constant field, the equations of motion for srings reduce to the usual one without $B_{\mu\nu}$. It is straightforward to include the effects of a constant B field by calculating the spectrum of the string. The inclusion of a constant $B_{\mu\nu}$ has some interesting consequences, one of these  is that there is a critical value which makes the energy of the winding modes aligned with $B_{\mu\nu}$ field vanish.

We had expected the modification induced by $B_{\mu\nu}$ to be significant since its presence makes the energy and the pressure of the winding modes vanish at a critical value. In fact our numerical results indicate that the behavior of the scale factor could be significantly affected.
The spacial directions expand like radiation dominated universe even with the presence of the both momentum and winding modes. 
%As the dilaton goes to weak coupling, the expansion of the universe stops and the solution for the scale factor becomes constant.

%Of course, this cannot be the whole story, it remains to see what could be the result when the two-form field is not constant and non-trivial effects could be expected when considering this case.

\section*{Acknowledgments}

\hspace{0.7cm}
This work is supported in part by Grant-in-Aid for Scientific Research from
the Ministry of Education, Science and Culture of Japan.

%%%%%
%%%%%%%%
%%%%%%%%%%%%%
%%%%%%%%%%%%%
%%%%%%%%%%%%%

\appendix

\section{Relation between string frame and Einstein frame}
\setcounter{equation}{0}
In this appendix, we summarize the relation between string frame and Einstein frame in our setup.
The parameter $d$ in this appendix which counts the number of the spacial dimensions should be put $d=9$ in superstring. 
From the conformal transformation (\ref{trans-string-einstein}) and the corresponding metrics the relation between scale factors is
\be
\alpha=-\frac{\phi}{d-1}+\lambda~,\quad  \beta=-\frac{\phi}{d-1}+\nu~
\ee
\be 
d\tilde{t}^2=e^{-\frac{2\phi}{d-1}}dt^2~
\ee
Also, the shifted dilaton is defined by
\be
\varphi \equiv \phi -\sum_{i=1}^{d} \lambda_i ~
\ee
Because of the presence of $H_{\mu\nu\lambda}$ for the superstring in ten dimensions, the spacial coordinates factorize as $T^3\times T^6$. Defining $\lambda_i\equiv\lambda$ for $i=\{1,2,3\}$ and $\lambda_j\equiv \nu$ for $j=\{4,\cdots ,9\}$, we have $\varphi = \phi - 3\lambda - 6\nu
$.  Using the Einstein equations and the solution for the homogeneous two-form field, we obtain the equations of motion for the superstring case $(d=9)$
\bea
\dot\varphi^2 -3\dot\lambda^2 -6\dot\nu^2 &=& e^\varphi E +U_o(\lambda)~ \\
\ddot\varphi - 3\dot\lambda^2 -6\dot\nu^2 &=&\frac{1}{2}e^\varphi E~\\
\ddot\lambda-\dot\varphi\dot\lambda &=&\frac{1}{2}P_\lambda +U_o(\lambda)~\\
\ddot\nu -\dot\varphi\dot\nu &=&\frac{1}{2}e^\varphi P_\nu~
\eea
with \be
U_o(\lambda)\equiv \frac{1}{12}H^2_{\alpha\beta\gamma}= \frac{1}{2}H_o^2 e^{-6\lambda}~
\ee
Accordingly, the equations of motion in the Einstein frame are 
\bea
6\alpha^{\prime 2}+ 36\alpha^\prime\beta^\prime +30\beta^{\prime 2}  &=&
\frac{\phi^{\prime 2}}{d-1}+e^{\frac{(d+1)\phi}{(d-1)}-3\alpha- 6\beta}E +\frac{1}{2}H_o^2 e^{{-6\alpha}-\frac{4\phi}{d-1}}~ \label{friedmann-einstein}\\
\alpha^{\prime\prime}+\alpha^\prime(3\alpha^\prime+6\beta^\prime)&=&
+\frac{3}{8}H_o^2 e^{-6\alpha-\frac{4\phi}{d-1}}+\frac{1}{2\cdot 8}e^{\frac{(d+1)}{(d-1)}\phi-3\alpha-6\beta}(-5E+5P_\lambda-6P_\nu)~ \label{Einstein-alpha} \\
\beta^{\prime\prime}+\beta^\prime(3\alpha^\prime+6\beta^\prime)&=&
-\frac{1}{8}H_o^2 e^{-6\alpha-\frac{4\phi}{d-1}}+\frac{1}{2\cdot 8}e^{\frac{(d+1)}{(d-1)}\phi-3\alpha-6\beta}(3E-3P_\lambda+2P_\nu)~\label{Einstein-beta}\\
\phi^{\prime\prime}+\phi^\prime (3\alpha^\prime +6\beta^\prime)&=&
H_o^2 e^{-6\alpha -\frac{4\phi}{d-1}}+\frac{1}{2}e^{\frac{(d+1)}{(d-1)}\phi-3\alpha -6\beta}(3P_\lambda + 6P_\nu)~
\eea
In the string frame, the equation of motion for $\lambda$ contains the dilaton and its time derivative but it does not contain $\nu$ terms. In the same way the equation of motion for $\nu$ is independent of $\lambda$ or its time derivatives. Then,  the equations of motion for the scale factors $\nu$ and $\lambda$ decouple and we can proceed to solve them numerically. In comparison, in the Einstein frame, the presence of $\tilde H_{\mu\nu\lambda}$ makes the scale factors  couple to each other. In spite of this unfavourable characteristic, the equations of motion in the Einstein frame are also useful, both  when trying to solve the equations of motion and also for clarifying the interpretation of the solutions. 

In the Einstein frame the field $\tilde H_{\mu\nu\lambda}$ is included in the equation of motion for both $\alpha$ and  $\beta$. By looking at the sign of the $H_o$ term in (\ref{Einstein-alpha}) and (\ref{Einstein-beta})  we can see  that the two-form field induces an anisotropic expansion on the scale factors, with $\phi$ and $\alpha$ being driven towards positive values while $\beta$ goes towards  negative values.  Also, while in the string frame it is possible to find solutions to the equations of motion in which $\nu$ becomes constant, this does not imply that the physical scale factor is fixed because it remains to stabilize the value of the dilaton. This can be seen directly from the relations of the Einstein frame to the string frame, where, in the case of $\nu=constant$ we have
\be
\beta = -\frac{\phi}{d-1} +{\text const}.
\ee
That is, unless both the dilaton $\phi$  and $\nu$ are constant in the string frame, there is no solution with $\beta=constant$ in the Einstein frame.

\newpage

%%%%%%%%%%%%%%%%%%%%%
%%%%%%%%%%%%%%%%%%%%%
%%%%%%%%%%%%%%%%%%%
%%%%%%%%%%%%%%%%%%
%%%%%%%%%%%%%%%%%%
%%%%%%%%%%%%%%%%%%


\begin{thebibliography}{99}
%%%%%%%%%%%%%%%%%%
%%%%%%%%%%%%%%%%%%
%%%%%%%%%%%%%%%%%%


%\cite{Brandenberger:1988aj}
\bibitem{Brandenberger:1988aj}
  R.~H.~Brandenberger and C.~Vafa,
  ``Superstrings in the Early Universe,''
  Nucl.\ Phys.\  B {\bf 316}, 391 (1989).
  %%CITATION = NUPHA,B316,391;%%
  
  %\cite{Brandenberger:2008nx}
\bibitem{Brandenberger:2008nx}
  R.~H.~Brandenberger,
  ``String Gas Cosmology,''
  arXiv:0808.0746 [hep-th].
  %%CITATION = ARXIV:0808.0746;%%

%\cite{Tseytlin:1991xk}
\bibitem{Tseytlin:1991xk}
  A.~A.~Tseytlin and C.~Vafa,
  ``Elements Of String Cosmology,''
  Nucl.\ Phys.\  B {\bf 372}, 443 (1992)
  [arXiv:hep-th/9109048].
  %%CITATION = NUPHA,B372,443;%%

%\cite{Tseytlin:1991ss}
\bibitem{Tseytlin:1991ss}
  A.~A.~Tseytlin,
  ``Dilaton, winding modes and cosmological solutions,''
  Class.\ Quant.\ Grav.\  {\bf 9}, 979 (1992)
  [arXiv:hep-th/9112004].
  %%CITATION = CQGRD,9,979;%%
%
%\cite{Battefeld:2005av}
\bibitem{Battefeld:2005av}
  T.~Battefeld and S.~Watson,
  ``String gas cosmology,''
  Rev.\ Mod.\ Phys.\  {\bf 78}, 435 (2006)
  [arXiv:hep-th/0510022].
  %%CITATION = RMPHA,78,435;%%
  
  %\cite{Sakellariadou:1995vk}
\bibitem{Sakellariadou:1995vk}
  M.~Sakellariadou,
  ``Numerical Experiments in String Cosmology,''
  Nucl.\ Phys.\  B {\bf 468}, 319 (1996)
  [arXiv:hep-th/9511075].
  %%CITATION = NUPHA,B468,319;%%
%  
%\cite{Campos:2003ip}

%\cite{Cleaver:1994bw}
\bibitem{Cleaver:1994bw}
  G.~B.~Cleaver and P.~J.~Rosenthal,
  ``String cosmology and the dimension of space-time,''
  Nucl.\ Phys.\  B {\bf 457}, 621 (1995)
  [arXiv:hep-th/9402088].
  %%CITATION = NUPHA,B457,621;%%
  
   %\cite{Bassett:2003ck}
\bibitem{Bassett:2003ck}
  B.~A.~Bassett, M.~Borunda, M.~Serone and S.~Tsujikawa,
  ``Aspects of string-gas cosmology at finite temperature,''
  Phys.\ Rev.\  D {\bf 67}, 123506 (2003)
  [arXiv:hep-th/0301180].
  %%CITATION = PHRVA,D67,123506;%%
  
  %\cite{Easson:2001fy}
\bibitem{Easson:2001fy}
  D.~A.~Easson,
  ``Brane gases on K3 and Calabi-Yau manifolds,''
  Int.\ J.\ Mod.\ Phys.\  A {\bf 18}, 4295 (2003)
  [arXiv:hep-th/0110225].
  %%CITATION = IMPAE,A18,4295;%%

%\cite{Greene:2009gp}
\bibitem{Greene:2009gp}
  B.~Greene, D.~Kabat and S.~Marnerides,
  ``Dynamical Decompactification and Three Large Dimensions,''
  arXiv:0908.0955 [hep-th].
  %%CITATION = ARXIV:0908.0955;%%
  
 
  
  \bibitem{Kitazawa:1998}
Hajime Aoki, Satoshi Iso, Hikaru Kawai, Yoshihisa Kitazawa, Tsukasa Tada, 
``Space-time structures from IIB matrix model,"
Prog.\ Theor.\ Phys.\ {\bf99}, 713 (1998). 

\bibitem{Kitazawa:2004}
Takaaki Imai, Yoshihisa Kitazawa, Yastoshi Takayama, Dan Tomino,
``Effective actions of matrix models on homogeneous spaces,"
Nucl.\ Phys.\ B{\bf 679}, 143 (2004). 

%\cite{Copeland:1994km}
\bibitem{Copeland:1994km}
  E.~J.~Copeland, A.~Lahiri and D.~Wands,
  ``String cosmology with a time dependent antisymmetric tensor potential,''
  Phys.\ Rev.\  D {\bf 51}, 1569 (1995)
  [arXiv:hep-th/9410136].
  %%CITATION = PHRVA,D51,1569;%%
  
  
\bibitem{Campos:2003ip}
  A.~Campos,
  ``Late cosmology of brane gases with a two-form field,''
  Phys.\ Lett.\  B {\bf 586}, 133 (2004)
  [arXiv:hep-th/0311144].
  %%CITATION = PHLTA,B586,133;%%
%
  %\cite{Campos:2005da}
\bibitem{Campos:2005da}
  A.~Campos,
  ``Asymptotic cosmological solutions for string / brane gases with  solitonic
  fluxes,''
  Phys.\ Rev.\  D {\bf 71}, 083510 (2005)
  [arXiv:hep-th/0501092].
  %%CITATION = PHRVA,D71,083510;%%

%\cite{Greene:2008hf}
\bibitem{Greene:2008hf}
  B.~Greene, D.~Kabat and S.~Marnerides,
  ``Bouncing and cyclic string gas cosmologies,''
  arXiv:0809.1704 [hep-th].
  %%CITATION = ARXIV:0809.1704;%%
%

%\cite{Goldwirth:1993ha}
\bibitem{Goldwirth:1993ha}
  D.~S.~Goldwirth and M.~J.~Perry,
  ``String Dominated Cosmology,''
  Phys.\ Rev.\  D {\bf 49}, 5019 (1994)
  [arXiv:hep-th/9308023].
  %%CITATION = PHRVA,D49,5019;%%


%\cite{Ambjorn:2000yr}
\bibitem{Ambjorn:2000yr}
  J.~Ambjorn, Y.~M.~Makeenko, G.~W.~Semenoff and R.~J.~Szabo,
  ``String theory in electromagnetic fields,''
  JHEP {\bf 0302}, 026 (2003)
  [arXiv:hep-th/0012092].
  %%CITATION = JHEPA,0302,026;%%
  
  
  %\cite{Grignani:2001ik}
\bibitem{Grignani:2001ik}
  G.~Grignani, M.~Orselli and G.~W.~Semenoff,
  ``The target space dependence of the Hagedorn temperature,''
  JHEP {\bf 0111}, 058 (2001)
  [arXiv:hep-th/0110152].
  %%CITATION = JHEPA,0111,058;%%

%\cite{Lidsey:1999mc}
\bibitem{Lidsey:1999mc}
  J.~E.~Lidsey, D.~Wands and E.~J.~Copeland,
  %``Superstring cosmology,''
  Phys.\ Rept.\  {\bf 337}, 343 (2000)
  [arXiv:hep-th/9909061].
  %%CITATION = PRPLC,337,343;%%

  
%%CITATION = JCAPA,0401,006;%%
\bibitem{WheelerGravitation}
 Charles Misner and Kip Thorne and John Wheeler,
  ``Gravitation'', ~(1973)


 %\cite{Easson:2001fy}
\bibitem{Freund-Rubin}
P. G. O. Freund and M. A. Rubin, ``Dynamics of dimensional reduction'', Phys. Lett. B
Volume 97, Issue 2, 1 December 1980, Pages 233-235.


%\cite{Alexander:2000xv}
\bibitem{Alexander:2000xv}
  S.~Alexander, R.~H.~Brandenberger and D.~Easson,
  ``Brane gases in the early universe,''
  Phys.\ Rev.\  D {\bf 62}, 103509 (2000)
  [arXiv:hep-th/0005212].
  %%CITATION = PHRVA,D62,103509;%%


%\cite{Easther:2002mi}
\bibitem{Easther:2002mi}
  R.~Easther, B.~R.~Greene and M.~G.~Jackson,
  ``Cosmological string gas on orbifolds,''
  Phys.\ Rev.\  D {\bf 66}, 023502 (2002)
  [arXiv:hep-th/0204099].
  %%CITATION = PHRVA,D66,023502;%%


%\cite{Easther:2003dd}
\bibitem{Easther:2003dd}
  R.~Easther, B.~R.~Greene, M.~G.~Jackson and D.~N.~Kabat,
  ``Brane gases in the early universe: Thermodynamics and cosmology,''
  JCAP {\bf 0401}, 006 (2004)
  [arXiv:hep-th/0307233].



%\cite{Easther:2004sd}
\bibitem{Easther:2004sd}
  R.~Easther, B.~R.~Greene, M.~G.~Jackson and D.~N.~Kabat,
``String windings in the early universe,''
  JCAP {\bf 0502}, 009 (2005)
  [arXiv:hep-th/0409121].
  %%CITATION = JCAPA,0502,009;%%

%\cite{Polchinski:1988cn}
\bibitem{Polchinski:1988cn}
  J.~Polchinski,
  ``Collision Of Macroscopic Fundamental Strings,''
  Phys.\ Lett.\  B {\bf 209}, 252 (1988).
  %%CITATION = PHLTA,B209,252;%%













  










%\cite{Polchinski:1995mt}
%\bibitem{Polchinski:1995mt}
%  J.~Polchinski,
 % ``Dirichlet-Branes and Ramond-Ramond Charges,''
 % Phys.\ Rev.\ Lett.\  {\bf 75}, 4724 (1995)
 % [arXiv:hep-th/9510017].
  %%CITATION = PRLTA,75,4724;%%

%\cite{Polchinski:1996na}
%\bibitem{Polchinski:1996na}
 % J.~Polchinski,
 %``Lectures on D-branes,''
 % arXiv:hep-th/9611050.
  %%CITATION = HEP-TH/9611050;%%

%\cite{Sen:1995cf}
%\bibitem{Sen:1995cf}
 % A.~Sen,
  %``T-Duality of p-Branes,''
 % Mod.\ Phys.\ Lett.\  A {\bf 11}, 827 (1996)
  %[arXiv:hep-th/9512203].
  %%CITATION = MPLAE,A11,827;%%




 

%\cite{Watson:2002nx}
\bibitem{Watson:2002nx}
  S.~Watson and R.~H.~Brandenberger,
  ``Isotropization in brane gas cosmology,''
  Phys.\ Rev.\  D {\bf 67}, 043510 (2003)
  [arXiv:hep-th/0207168].
  %%CITATION = PHRVA,D67,043510;%%

%\cite{Easther:2002qk}
%\bibitem{Easther:2002qk}
%  R.~Easther, B.~R.~Greene, M.~G.~Jackson and D.~N.~Kabat,
%  ``Brane gas cosmology in M-theory: Late time behavior,''
 % Phys.\ Rev.\  D {\bf 67}, 123501 (2003)
%  [arXiv:hep-th/0211124].
  %%CITATION = PHRVA,D67,123501;%%


 



%\cite{Brandenberger:2001kj}
%\bibitem{Brandenberger:2001kj}
 % R.~Brandenberger, D.~A.~Easson and D.~Kimberly,
  %``Loitering phase in brane gas cosmology,''
 % Nucl.\ Phys.\  B {\bf 623}, 421 (2002)
%  [arXiv:hep-th/0109165].
  %%CITATION = NUPHA,B623,421;%%


%\cite{Maggiore:1998cz}
%\bibitem{Maggiore:1998cz}
  %M.~Maggiore and A.~Riotto,
  %``D-branes and cosmology,''
  %Nucl.\ Phys.\  B {\bf 548}, 427 (1999)
  %[arXiv:hep-th/9811089].
  %%CITATION = NUPHA,B548,427;%%
  
  




  
    
\end{thebibliography}
\end{document}